\documentclass[11pt]{article}

\usepackage[title]{appendix} % The 'title' option adds the word 'Appendix'

\usepackage[margin=1in]{geometry}
\usepackage[T1]{fontenc}
\usepackage[utf8]{inputenc}
\usepackage{lmodern}
\usepackage{microtype}
\usepackage{hyphenat}
\usepackage{setspace}

\usepackage{amsmath,amssymb}
\usepackage{graphicx}
\usepackage{authblk}
\usepackage{booktabs,array,tabularx,longtable}
\usepackage[table,dvipsnames]{xcolor}
\usepackage{caption}
\usepackage{subcaption}
\usepackage{float}
\usepackage{adjustbox}
\usepackage{makecell}

\usepackage{enumitem}
\usepackage[round,authoryear]{natbib}
\usepackage[normalem]{ulem}

\usepackage[utf8]{inputenc}
\usepackage{graphicx,epsfig}
\usepackage{epstopdf}
\usepackage{amssymb,amsmath,amsthm,amsfonts}
\usepackage{bm}
\usepackage[dvipsnames]{xcolor}
\usepackage[linktocpage,breaklinks]{hyperref}
\hypersetup{colorlinks=true,
            citecolor=NavyBlue,
            linkcolor=NavyBlue,
            urlcolor=NavyBlue}
\usepackage{url}
\usepackage{xurl} % allows better line breaks in paths/URLs

\graphicspath{{figures/}}

\definecolor{lightaccent}{RGB}{236,242,248}

\newcommand{\Ht}{H_t}
\newcommand{\phit}{\phi_t}
\newcommand{\Dt}{D_t}
\newcommand{\csvfile}[1]{\nolinkurl{#1}}
\newcommand{\surface}{\textit{surface}}
\newcommand{\midstream}{\textit{mid-stream}}
\newcommand{\residue}{\textit{residue}}

\title{Measuring Collective Semantic Change in Populations of Language Model Agents}

\author{Elena Kopteva\thanks{\texttt{koptieva@illinois.edu}}}
\affil{The Grainger College of Engineering,
Department of Physics,\\ University of Illinois Urbana-Champaign, Urbana, Illinois 61801, USA}

\date{August 2026}

\begin{document}

\maketitle

\begin{abstract}
Collective semantic change in populations of language model agents is a measurable dynamical phenomenon. We present a passive longitudinal instrument called Kopterix that observes the semantic state of an agent population as a sequence of bounded observations under a protocol defined before the observations begin. Each observation divides the sampled feed by post age into surface, mid-stream, and residue layers, which makes semantic differences across content age measurable alongside run-to-run change. We validate the instrument on Moltbook, an agent-native social platform, over a two-month window of scheduled observations, with the periodicity check extended across approximately four months. At the lexical level, rarefied entropy resolves an April-May difference in the evenness of the stored top 200 unigram distributions, and adjacent states are lexically closer than states paired after timestamp shuffling. At the geometric level, grand mean centering exposes the scale of a common embedding direction, and scheduled shuffle checks support a recurring excess in the mid-stream to residue separation relative to the shuffled reference. At the temporal level, detrended scalar quantities and centered layer centroids lose much of their similarity over several hours, and a weaker positive component declines across longer separations with no strong weekly recurrence. Several attractive apparent structures failed their controls, and each reading is limited to the level its controls support. The design applies wherever a population of agents produces a timestamped language environment that can be observed repeatedly and divided by content age.
\end{abstract}

\newpage

\tableofcontents

%%%%%%%%%%%%%%%%%%%%%%%%%%%%%%%%%%%%%%%
%%%%%%%%%%%%%%%%%%%%%%%%%%%%%%%%%%%%%%%

\section{Introduction}
\label{sec:introduction-motivation}

Agent-native social platforms, on which autonomous language model agents post and interact with little or no direct human participation, became a concrete empirical setting in early 2026, with Moltbook as the central case studied so far
\citep{demarzo2026collective,feng2026moltnet,jiang2026humans,goyal2026socialsimulacra,hou2026structural}.
These platforms provide an observable setting in which populations of LLM-based agents produce, read, and respond through a shared textual environment over time. They also bring into view a broader problem as LLM-based reasoning agents begin to work as autonomous teams over long tasks and missions.

Coordination in such teams depends on language, interpreted context, and evolving representations of goals. Meaning becomes part of the multi-agent system's working medium. Individual agents may continue to respond, complete tasks, and report healthy status while the semantic organization of the team is changing. As subgroups emerge, interpretations of context may diverge, priorities may drift, and the shared representation of the mission may lose coherence before the system exhibits a conventional failure. Ordinary system monitoring does not directly measure this process.

Early studies already report related collective effects and coordination failures in LLM-based agent teams. Analyses of execution traces identify inter-agent misalignment through task derailment, information withholding, ignored agent input, and mismatches between reasoning and action \citep{cemri2025multiagentfailures}. In teams that coordinate through discussion, consensus seeking can dilute expert information and leave the group below the performance of its strongest member \citep{pappu2026experts}. Controlled populations also develop shared conventions and collective biases that are absent from individual behavior, with interaction and group size shaping the resulting collective states \citep{ashery2025conventions,flint2025groupsize}. Together, these results show that collective organization can change while agents continue to interact and generate outputs. These studies use task failure, team performance, or collective outcomes in controlled experiments as their primary endpoints. They leave open whether changes in collective semantic organization can be measured as a temporal process before a terminal outcome is observed.

This leads to two connected questions.

The practical question is whether collective semantic change can be measured early enough to support mission assurance in autonomous teams.

The broader research question concerns the collective semantic dynamics of populations of LLM-based agents. Their outputs are generated by algorithmic systems, and their communication passes through shared textual environments. These conditions may produce collective dynamics with no direct human analogue. Their structure must be measured directly to be understood. The relevant phenomena include semantic drift, persistence, regime shifts, recurrence, fragmentation, and emerging structure.

This work introduces a standing measurement instrument for repeated observation of collective semantic change in an open agent-native population. Each run stratifies one feed observation by post age and compares the resulting semantic layers within the run. The measurements are interpreted under stated support conditions, internal controls, and a defined inference ceiling. This reading discipline is part of the instrument itself. We test whether the design resolves lexical, embedding, geometric, and temporal structure in the sampled population and whether relations among the measured quantities support controlled readings of collective semantic change within the inference ceiling of the available record.

We implement the measurement design in a software instrument called Kopterix and apply it to Moltbook over the half-open UTC interval $[2026\text{-}04\text{-}01, 2026\text{-}06\text{-}01)$ under a fixed sampling protocol with three layers defined by post age. This paper presents the instrument design and its main measurements, reports how those measurements respond to changes in the sampled Moltbook content under internal controls, and identifies where interpretation remains unresolved. The periodicity check uses observations continued through early August, which extends that part of the record to approximately four months. The complete two-month validation report, archived data products, analysis code, and provenance record are available in the accompanying Zenodo deposit \citep{kopterix2026validation}.

Moltbook serves as the validation setting for this design. Operational agent teams remain a later application domain. Because the Moltbook feed is currently retrievable as a complete archive, it also supports retrospective replay and exhaustive corpus analysis of the platform. That is a different empirical question from the standing observation problem addressed here.

The standing instrument is designed to measure semantic change as it occurs in agent populations whose history may be incomplete, whose content may be only partly accessible, or whose model and platform conditions may change. Its value lies in defining the observation schedule, measurements, and control baselines before observations begin. This remains important even when a corpus can later be stored, because the relevant time scale of change is itself an empirical quantity and assurance tasks require deviations to be detected while the multi-agent system is still operating. In the present study, the recoverability of the Moltbook feed provides a reference record against which the observational protocol can be checked before the design is applied to systems in which retrospective replay is unavailable.

Within the two-month observation period, the instrument resolves changes in lexical concentration, relations between within-run homogeneity and run-to-run drift, cross-layer geometry, and temporal dependence. Together, these measurements support controlled readings of semantic drift and temporal persistence and permit the collective semantic state of a multi-agent system to be studied through the geometric vocabulary for dynamical systems developed by \citet{tacheny2026geometric}.

%%%%%%%%%%%%%%%%%%%%%%%%%%%%%%%%%%%%%%%
%%%%%%%%%%%%%%%%%%%%%%%%%%%%%%%%%%%%%%%

\section{Relation to existing work}
\label{sec:relation-to-work}

Although research in this area is still emerging, several distinct directions are already visible in the literature.

One direction treats Moltbook as an interaction graph, with agents as nodes and comments or replies as edges. These studies examine connectivity, central actors, community structure, fragility, and coordination across the platform \citep{feng2026moltnet,hou2026structural,sodano2026emergencefragilityllmbasedsocial,demarzo2026collective,mukherjee2026moltgraph}. MoltGraph extends this approach across time by following changes in the interaction graph over $30$ days of repeated observations \citep{mukherjee2026moltgraph}. The present study follows how the collective semantic state of the feed changes over time through repeated measurements of sampled text, including differences among post age layers within each observation run.

A second direction works with large stored corpora of Moltbook posts and comments. These studies examine topics, toxicity, social structure, lexical properties, and semantic geometry \citep{jiang2026humans,goyal2026socialsimulacra,demarzo2026collective,brach2026moltbookfiles}. The Moltbook Observatory Archive supports longitudinal analysis by providing a record partitioned by date and assembled through continuous polling \citep{gautam2026observatory}. Work at this scale also shows that the feed contains a large transactional substrate alongside natural language discourse, and that reply volume alone gives limited evidence of substantive interaction \citep{ayan2026platform,shekkizhar2026interaction}. Such archives support large scale retrospective analysis over their stored records. They also make the mixed composition of the feed an explicit interpretation limit for the present measurements.

Controlled experiments have already shown that LLM agent groups can develop recognizable forms of collective behavior. They can form conventions and collective biases, and can also fail systematically to surface information distributed across the group during team reasoning \citep{ashery2025conventions,li2025collectivefailures}. These results show that collective organization is itself a measurable property of agent populations. In these studies, the interaction rules, tasks, and outcomes are experimentally specified. The present study brings this measurement problem into an open population whose composition, prompts, exposure, and objectives are not experimentally controlled.

The measurement strategy used here is closest to work that traces semantic development through repeated observations and represents change in embedding space. \citet{li2026socialization} measures semantic stabilization, lexical turnover, individual inertia, and influence persistence in a retrospective Moltbook corpus. It finds rapid global stabilization with little evidence of progressive adaptation and identifies shared temporal context as a rival explanation for apparent semantic influence. \citet{li2026attraction} uses the same sentence embedding model that we use in this study. The authors follow weekly mean pairwise cosine similarity within communities and compare the observed convergence slopes with temporal permutation baselines. They find a declining global baseline across the selected communities, while some communities become more internally similar. Their stable cohort analysis links this local convergence to selective attraction and differential retention. \citet{tacheny2026geometric} provides a geometric vocabulary of trajectories and regimes for recursive transformations in closed agent loops. Together, these studies show how semantic change can be traced through temporal diagnostics, controlled comparisons, and geometric descriptions. Their measured objects are full corpora, communities, individual agents, or closed recursive trajectories.

We found no prior measurement design for an open online agent-native population that divides each feed observation by content age and compares the resulting semantic layers within the same run. This leaves a practical gap in measuring collective semantic change as it occurs under an observational protocol whose schedule, measurements, and control baselines are defined before observations begin. The present study addresses this gap with a standing instrument and tests what can be inferred about ongoing change in collective semantic state under explicit support conditions and internal controls.

%%%%%%%%%%%%%%%%%%%%%%%%%%%%%%%%%%%%%%%
%%%%%%%%%%%%%%%%%%%%%%%%%%%%%%%%%%%%%%%

\section{Instrument design}\label{sec:instrument-design}

A single Kopterix run is one observation of a bounded feed sample, and the run as a whole is the unit of interpretation. This section describes the standing design of an observation, defines the state that each observation records, and outlines how far the resulting quantities may be read. Kopterix is the instrument prototype, and Moltbook is the setting in which it is presently validated.

%===================================

\subsection{Observation protocol}
\label{sec:observation-protocol}

An observation begins with a fetch. The instrument draws posts from Moltbook in three groups defined by how long ago a post appeared, which we call post age layers. The \surface{} layer holds posts from the last $0$-$2$ hours, the \midstream{} layer holds posts from $3$-$8$ hours before the run, and the \residue{} layer holds posts from $12$-$24$ hours before the run. Together the three layers give one run a shallow view backward in time, so that recent material and older material can be compared inside a single sample. Each layer has a hard cap of $250$ fetched posts, which places at most $750$ posts in a run with all three layers present.

Only part of each post enters the measurement. For every fetched post, the measured text, which we call the stored text, is its title followed by a truncated prefix of its body,
\begin{equation*}
\text{stored text} = \text{title} + \text{first 300 characters of content}.
\end{equation*}
Full post bodies stay outside the measurement. Each stored text is a bounded textual projection of the post, and the bound is fixed in advance for every fetched post.

From these stored texts the deterministic measurements follow. The instrument forms token counts, sentence embeddings, run mean embeddings, layer centroids, and the state needed to compare the present run with the preceding one.

The instrument records two row types. The observation row holds the timestamp, the sample support, the run-time scalar measurements, and the provenance fields. The state row holds the run mean embedding, the layer centroids, and the top $200$ unigram counts used for later lexical comparison.

%===================================

\subsection{Measured quantities}
\label{sec:measured-quantities}

The deterministic measurements use four inputs. These are token counts from lowercased stored texts split on whitespace, sentence embeddings of usable stored texts, the post age layer assigned to each fetched post, and the state stored by the preceding run.

A stored text is usable for embedding analysis when it is nonempty. The embedding measurements for a run are skipped when fewer than $10$ usable stored texts remain.

Let $t$ index runs, let $i$ index usable stored texts within run $t$, and let $\ell\in\{S,M,R\}$ index the post age layers. Here, $S$, $M$, and $R$ denote the surface, mid-stream, and residue layers.

To compare stored texts by semantic content, we map each usable stored text $i$ in run $t$ to a vector
\begin{equation*}
e_i(t)\in\mathbb{R}^{384}
\end{equation*}
using \csvfile{sentence-transformers/all-MiniLM-L6-v2}
\citep{reimers2019sentencebert,allminilm}.
The vector $e_i(t)$ is the sentence embedding of the stored text. Under this model, texts with similar semantic content tend to occupy nearby positions in the embedding space. Their similarity and displacement can then be evaluated with vector operations. Comparisons between consecutive runs use the stored state from the preceding run.

\paragraph{Sample support and lexical state.}

Before any comparison, we record the number of fetched posts in each layer and in the full run. These counts define the sample support,
\begin{equation*}
n_{\mathrm{surface}}(t),\quad
n_{\mathrm{mid}}(t),\quad
n_{\mathrm{residue}}(t),\quad
n_{\mathrm{total}}(t).   
\end{equation*}
These counts refer to fetched posts. The sample support is read before any scalar or layer comparison because several quantities below depend on sample size.

Lexical entropy describes how evenly token mass is spread over the observed vocabulary. Let $p_t(w)$ be the empirical frequency of token $w$ across all stored texts in run $t$. Then
\begin{equation}
\Ht = -\sum_w p_t(w)\log_2 p_t(w).
\end{equation}
Higher $\Ht$ indicates that token mass is distributed more broadly across the observed token types, and lower $\Ht$ indicates greater lexical concentration. Raw $\Ht$ is sensitive to sample support. For this reason, it must be read together with $n_{\mathrm{total}}$.

To compare lexical entropy across runs with different sample support, we use rarefaction at a common token budget. For each run $t$, the unigram count of token $w$ is the number of times that token occurs across all stored texts. The instrument stores the counts of the $200$ most frequent tokens and normalizes them to form a truncated empirical token distribution. Let $p_j^{(B)}(w)$ be the empirical frequency of token $w$ in draw $j$ of $B$ tokens from this distribution. Averaging the entropy over $J$ draws gives rarefied entropy,
\begin{equation}
H_{\mathrm{rare}}(t;B)
=
\frac{1}{J}
\sum_{j=1}^{J}
\left[
-\sum_w p_j^{(B)}(w)\log_2 p_j^{(B)}(w)
\right].
\end{equation}
Because the draws are restricted to the stored top $200$ unigram counts, $H_{\mathrm{rare}}(t;B)$ measures lexical evenness within that truncated distribution.

\paragraph{Run embedding quantities.}

The embedding quantities describe the central position of the embedded texts in a run, their average similarity, and the movement of that central position between consecutive runs. Let $N_t$ be the number of usable stored texts with embeddings in run $t$. The run mean embedding is the average position of the embedded texts in run $t$,
\begin{equation}
\bar e(t)=\frac{1}{N_t}\sum_{i=1}^{N_t}e_i(t).
\end{equation}

Semantic homogeneity measures how similar the texts of a single run are to one another. It is the mean pairwise cosine similarity within the run,
\begin{equation}
\phit
=
\frac{2}{N_t(N_t-1)}
\sum_{i<j}
\cos\left(e_i(t),e_j(t)\right).
\end{equation}
Higher $\phit$ indicates greater average similarity among the embedded texts. The quantity is read as the degree of semantic homogeneity within the sampled run.

To measure change in the central position between consecutive runs, we use run-to-run drift. Let $\bar e(t^-)$ be the mean embedding stored by the preceding run. Run-to-run drift is
\begin{equation}
\Dt
=
1-\cos\left(\bar e(t),\bar e(t^-)\right).
\end{equation}
Low $\Dt$ indicates that the run mean has stayed close to its preceding value, and high $\Dt$ indicates greater displacement of that mean.

\paragraph{Layer state and residual geometry.}

The layer quantities compare the three post age strata inside a single run. For layer $\ell\in\{S,M,R\}$, let $\mathcal I_\ell(t)$ be the set of indices of usable embedded texts in that layer, and let $N_{\ell,t}=|\mathcal I_\ell(t)|$ denote the cardinality of this set, meaning the number of such texts. For $N_{\ell,t}>0$, the layer centroid is the average position of the embedded texts in layer $\ell$,
\begin{equation}
c_\ell(t)
=
\frac{1}{N_{\ell,t}}
\sum_{i\in\mathcal I_\ell(t)}
e_i(t).
\end{equation}

Comparing the centroids with one another gives the three cross-layer distances,
\begin{align}
d_{SM}(t) &= 1-\cos\left(c_S(t),c_M(t)\right),\\
d_{MR}(t) &= 1-\cos\left(c_M(t),c_R(t)\right),\\
d_{SR}(t) &= 1-\cos\left(c_S(t),c_R(t)\right).
\end{align}
A small distance indicates similar semantic centers for the two age strata.

To measure how far each layer centroid lies from the run mean embedding, we use the layer residual,
\begin{equation}\label{eq:layer-resid}
r_\ell(t)=c_\ell(t)-\bar e(t).
\end{equation}
The vector $r_\ell(t)$ gives the direction and magnitude of the displacement of layer $\ell$ from the run mean.

To compare the three residuals jointly, let
\begin{equation}\label{eq:equal-weight-mean}
\bar c(t)
=
\frac{1}{3}
\left(
c_S(t)+c_M(t)+c_R(t)
\right)
\end{equation}
denote the equal weight mean of the three layer centroids. The residuals then satisfy
\begin{equation}\label{eq:layer-residual-sum}
r_S(t)+r_M(t)+r_R(t)
=
3\left(\bar c(t)-\bar e(t)\right).
\end{equation}
Their sum is zero when the run mean embedding equals the equal weight mean of the layer centroids.

To place each run mean relative to the common center of the observed runs, we define the grand mean embedding across $T$ state rows,
\begin{equation}
\bar e_{\mathrm{grand}}
=
\frac{1}{T}
\sum_{t=1}^{T}\bar e(t).
\end{equation}
The grand mean residual is then
\begin{equation}
\delta e(t)
=
\bar e(t)-\bar e_{\mathrm{grand}}.
\end{equation}
The vector $\delta e(t)$ gives the displacement of the run mean embedding at run $t$ from the grand mean embedding.

\paragraph{Lexical and temporal change.}

Lexical change between consecutive runs is measured from the stored top $200$ unigram counts. Let $u(t)$ denote the stored unigram count vector for run $t$, with its components indexed by token. Before comparison, its components are aligned by token with those of $u(t^-)$, the vector stored by the preceding run. The consecutive lexical cosine step is the cosine distance between these two vectors,
\begin{equation}
d_{\mathrm{lex}}(t)
=
1-\cos\left(u(t),u(t^-)\right).
\end{equation}
Low values indicate that the stored unigram count vectors of consecutive runs resemble one another.

Temporal quantities describe dependence across runs separated by multiple observation steps. For a scalar series $x(t)$, the autocorrelation at lag $k$ is the Pearson correlation between values separated by $k$ observation steps, computed over pairs for which both values are available. Lag $k$ corresponds to an elapsed time determined by the observation schedule. For layer $\ell$, the layer centroid similarity at lag $k$, denoted by $\rho_{k,\ell}$, is the mean cosine similarity between layer centroids separated by $k$ observation steps. Layer centroid similarity uses cosine similarity. Scalar autocorrelation uses Pearson correlation.

Table~\ref{tab:core-vocabulary} summarizes the quantities defined above, their notation, and what each quantity measures. We refer collectively to $\Ht$, $H_{\mathrm{rare}}$, $\phit$, $\Dt$, $n_{\mathrm{total}}$, $d_{SM}$, $d_{MR}$, $d_{SR}$, and \texttt{post\_rate\_est} as the scalar measurements, where \texttt{post\_rate\_est} is the auxiliary estimate of recent posting activity.

\begin{center}
\renewcommand{\arraystretch}{1.5}
\small
\begin{longtable}{
p{0.30\textwidth}
p{0.12\textwidth}
p{0.56\textwidth}
}
\caption{Core quantities and notation.
\label{tab:core-vocabulary}}\\
\hline
\textbf{Quantity}
&
\textbf{Notation}
&
\textbf{What the quantity measures}
\\
\hline
\endfirsthead

\hline
\textbf{Quantity}
&
\textbf{Notation}
&
\textbf{What the quantity measures}
\\
\hline
\endhead

\hline
\endfoot

\hline
\endlastfoot

\multicolumn{3}{c}{\textit{Sample support}}
\\
\hline
\noalign{\smallskip}

Surface sample count
&
$n_{\mathrm{surface}}(t)$
&
Number of fetched posts in the surface layer of run $t$.
\\

Mid-stream sample count
&
$n_{\mathrm{mid}}(t)$
&
Number of fetched posts in the mid-stream layer of run $t$.
\\

Residue sample count
&
$n_{\mathrm{residue}}(t)$
&
Number of fetched posts in the residue layer of run $t$.
\\

Total sample count
&
$n_{\mathrm{total}}(t)$
&
Total number of fetched posts in run $t$.
\\

\noalign{\smallskip}
\hline
\multicolumn{3}{c}{\textit{Lexical quantities}}
\\
\hline
\noalign{\smallskip}

Stored unigram count vector
&
$u(t)$
&
Vector of the stored top $200$ unigram counts for run $t$, with components indexed by token.
\\

Lexical entropy
&
$\Ht$
&
Lexical evenness of the empirical token distribution in run $t$.
\\

Rarefied entropy
&
$H_{\mathrm{rare}}(t;B)$
&
Lexical evenness within the stored top $200$ unigram distribution at a common token budget $B$.
\\

Consecutive lexical cosine step
&
$d_{\mathrm{lex}}(t)$
&
Cosine distance between the stored unigram count vectors of run $t$ and the preceding run.
\\

\noalign{\smallskip}
\hline
\multicolumn{3}{c}{\textit{Run embedding quantities}}
\\
\hline
\noalign{\smallskip}

Run mean embedding
&
$\bar e(t)$
&
Mean position of the usable stored text embeddings in run $t$.
\\

Semantic homogeneity
&
$\phit$
&
Mean pairwise cosine similarity among the usable stored text embeddings in run $t$.
\\

Run-to-run drift
&
$\Dt$
&
Cosine distance between the run mean embedding at $t$ and the mean embedding stored by the preceding run.
\\

\noalign{\smallskip}
\hline
\multicolumn{3}{c}{\textit{Layer quantities}}
\\
\hline
\noalign{\smallskip}

Layer centroid
&
$c_\ell(t)$
&
Mean position of the usable stored text embeddings in layer $\ell$ of run $t$.
\\

Surface-mid distance
&
$d_{SM}(t)$
&
Cosine distance between the surface and mid-stream layer centroids in run $t$.
\\

Mid-residue distance
&
$d_{MR}(t)$
&
Cosine distance between the mid-stream and residue layer centroids in run $t$.
\\

Surface-residue distance
&
$d_{SR}(t)$
&
Cosine distance between the surface and residue layer centroids in run $t$.
\\

\noalign{\smallskip}
\hline
\multicolumn{3}{c}{\textit{Layer centering}}
\\
\hline
\noalign{\smallskip}

Layer residual
&
$r_\ell(t)$
&
Displacement of the centroid of layer $\ell$ from the run mean embedding.
\\

Equal weight layer mean
&
$\bar c(t)$
&
Mean of the three layer centroids with equal weight assigned to each layer.
\\

\noalign{\smallskip}
\hline
\multicolumn{3}{c}{\textit{Grand mean centering}}
\\
\hline
\noalign{\smallskip}

Grand mean embedding
&
$\bar e_{\mathrm{grand}}$
&
Mean position of the run mean embeddings across the included state rows.
\\

Grand mean residual
&
$\delta e(t)$
&
Displacement of the run mean embedding at $t$ from the grand mean embedding.
\\

\noalign{\smallskip}
\hline
\multicolumn{3}{c}{\textit{Temporal quantities}}
\\
\hline
\noalign{\smallskip}

Scalar autocorrelation at lag $k$
&
---
&
Pearson correlation between available values of a scalar series separated by $k$ observation steps.
\\

Layer centroid similarity at lag $k$
&
$\rho_{k,\ell}$
&
Mean cosine similarity between centroids of layer $\ell$ separated by $k$ observation steps.
\\

\end{longtable}
\end{center}

%===================================

\subsection{Reading the measurements}\label{sec:reading-measurements}
%\label{sec:interpretive-scope}

The quantities defined above are read under stated support conditions that set the epistemic ceiling for each interpretation. This subsection gives those conditions in the order in which the measurements are read.

Sample support is read first. A decrease in raw lexical entropy $\Ht$ supports a reading of greater lexical concentration only when $n_{\mathrm{total}}$ is stable across the runs under comparison or when rarefied entropy $H_{\mathrm{rare}}(t;B)$ changes in the same direction at a common token budget $B$. Without either condition, the decrease in raw $\Ht$ may reflect variation in sample support.

Semantic homogeneity $\phit$ and run-to-run drift $\Dt$ are read jointly to relate similarity within a run to movement between consecutive runs. When $\phit$ increases and $\Dt$ decreases, the embedded posts become more homogeneous while the run mean remains close to the preceding stored run mean. This pattern is compatible with local tightening, in which the sample becomes more coherent within the same semantic neighborhood. When both $\phit$ and $\Dt$ increase, the embedded posts become more homogeneous while the run mean moves away from the preceding stored run mean. This pattern is compatible with a coherent shift, in which the sample tightens while moving toward a different semantic neighborhood. 

The three cross-layer distances are read jointly as the relative geometry of the layer centroids within one run. Small values of $d_{SM}(t)$, $d_{MR}(t)$, and $d_{SR}(t)$ indicate nearby semantic centers across the three post age layers. Unequal values indicate that the three layer pairs differ in their degree of separation. These configurations describe the relative positions of the layer centroids and define specific geometric patterns for later tests. Observed message paths would allow tests of propagation or relay. Controlled interventions would allow tests of influence. Known subgroup labels would allow tests of subgroup structure.

A low value of the consecutive lexical cosine step $d_{\mathrm{lex}}(t)$ indicates that the stored unigram count vectors of consecutive runs are similar. Positive scalar autocorrelation at short lags indicates temporal dependence between values separated by a small number of observation steps. High layer centroid similarity $\rho_{k,\ell}$ indicates that layer centroids separated by $k$ observation steps remain similar on average. When a decay model is fitted to a temporal dependence curve, the fitted decay time summarizes the time scale over which the measured dependence decreases under that model.

The sampled Moltbook feed contains mechanically repeated minting and promotional posts alongside discursive posts. Every primary deterministic measurement is computed on this mixed feed. Therefore, changes in $\Ht$, $H_{\mathrm{rare}}$, $\phit$, $\Dt$, the layer geometry, and the temporal quantities describe the mixed feed as sampled.
Filtering mechanically repeated minting and promotional posts before measurement would allow the measured changes to be interpreted within the discursive component of the feed.

The stored top $200$ unigram counts include those minting tokens. The lexical sensitivity analysis recomputes $H_{\mathrm{rare}}$ after subtracting the identified mint payload tokens from the stored top $200$ unigram counts, as reported in Section~\ref{sec:mint-payload-sensitivity}. The embedding record stores only run and layer aggregates. Recomputing the embedding quantities on a discursive subset would require embeddings for the individual stored texts. Including these embeddings is an objective for future records.

To interpret temporal and layer structure, we compare the observed measurements with three internal control baselines in which the relevant organization of the data is altered. The deterministic shuffle comparison places the observed cross-layer distances beside distances computed from structured shuffled samples, providing a baseline for the amount of cross-layer separation produced by the shuffle procedure. The timestamp permutation baseline preserves the stored unigram count vectors and permutes their temporal order, providing a baseline for lexical similarity between consecutive runs. The exact layer label permutation compares the observed ordering of $d_{SM}$, $d_{MR}$, and $d_{SR}$ against all six assignments of the $S$, $M$, and $R$ labels, providing a baseline for whether that ordering is associated with the original post age labels. Together, these controls allow us to assess whether the observed separation, lexical adjacency, and layer ordering differ from their corresponding baselines.

Together, the measured quantities and the internal controls form a measurement vocabulary for collective semantic dynamics across runs, post age layers, and lags. The following analysis uses this vocabulary to distinguish changes in lexical concentration, relations between within-run homogeneity and run-to-run drift, cross-layer geometry, and temporal dependence in the sampled feed. These measurements also define the conditions under which collective semantic memory, recurrence, fragmentation, and regime change can be tested. Causal mechanisms require observed message paths or controlled interventions. Subgroup formation requires known subgroup labels. Behavior in operational agent teams requires observations from those systems and remains within the scope of our future work.

%%%%%%%%%%%%%%%%%%%%%%%%%%%%%%%%%%%%%%%
%%%%%%%%%%%%%%%%%%%%%%%%%%%%%%%%%%%%%%%

\section{Observation record}\label{sec:observation-record}

The observation window is the half-open UTC interval $[2026\text{-}04\text{-}01, 2026\text{-}06\text{-}01)$, from 2026-04-01 00:00 inclusive to 2026-06-01 00:00 exclusive. The observation record contains $231$ observation rows in \csvfile{observations.csv} and $215$ state rows in \csvfile{kopterix_state.csv}, with at least one observation on each day. The median gap between consecutive observations is $5.75$ hours, and the maximum gap is $23.64$ hours. For the full centroid temporal record, $210$ rows have complete layer centroids, and the median separation is $5.74$ hours.

The observation record contains $121$ rows in April and $110$ rows in May. A timestamp matching procedure linked $214$ of the $231$ observation rows to state rows. This left $17$ unmatched observation rows and $1$ unmatched state row.

Table~\ref{tab:observation-state-coverage} summarizes observation and state row coverage within the observation window.

\begin{table}[htbp]
\centering
\small
\setlength{\tabcolsep}{4pt}
\caption{Observation and state row coverage.}
\label{tab:observation-state-coverage}
\begin{tabular}{@{}llllll@{}}
\toprule
\makecell[l]{Observation\\rows in window} &
\makecell[l]{State rows\\in window} &
\makecell[l]{Observations\\matched to state} &
\makecell[l]{Unmatched\\observations} &
\makecell[l]{Unmatched\\state rows} &
\makecell[l]{Match\\tolerance (min)} \\
\midrule
231 & 215 & 214 & 17 & 1 & 5 \\
\bottomrule
\end{tabular}
\end{table}

The $231$ observation rows define the overall observation coverage. Individual analyses may use fewer rows when required fields are missing or when the fetched sample does not provide the support required for a particular quantity. The usable row count is reported separately for each analysis.

%%%%%%%%%%%%%%%%%%%%%%%%%%%%%%%%%%%%%%%
%%%%%%%%%%%%%%%%%%%%%%%%%%%%%%%%%%%%%%%

\section{Lexical measurements}
\label{sec:lexical-measurements}

Raw lexical entropy, $\Ht$, was initially used to compare lexical dispersion across runs. The first diagnostic showed that $\Ht$ depends strongly on sample support when $n_{\mathrm{total}}$ varies. This dependence sets the starting condition for the analysis in this section.

Here, we first measure the dependence of raw $\Ht$ on $n_{\mathrm{total}}$. Then  we recompute lexical entropy at a common token budget through rarefaction and use the adjusted quantity to compare the lexical distributions observed in April and May. Finally, we test the result against the six May observations with low $n_{\mathrm{total}}$ and the identified mint payload tokens.

%===================================

\subsection{Dependence of raw entropy on sample support}

To quantify the relation between raw lexical entropy, $\Ht$, and sample support, $n_{\mathrm{total}}$, we compute the Pearson correlation between the two quantities \citep{pearson1895note}. This correlation can be evaluated only when $n_{\mathrm{total}}$ varies across observations. The corresponding $p$ value indicates how surprising a correlation this far from zero would be if no linear association were present. Smaller $p$ values provide stronger evidence that the association is not zero.

\begin{figure}[ht]
\centering
\includegraphics[width=0.94\textwidth]{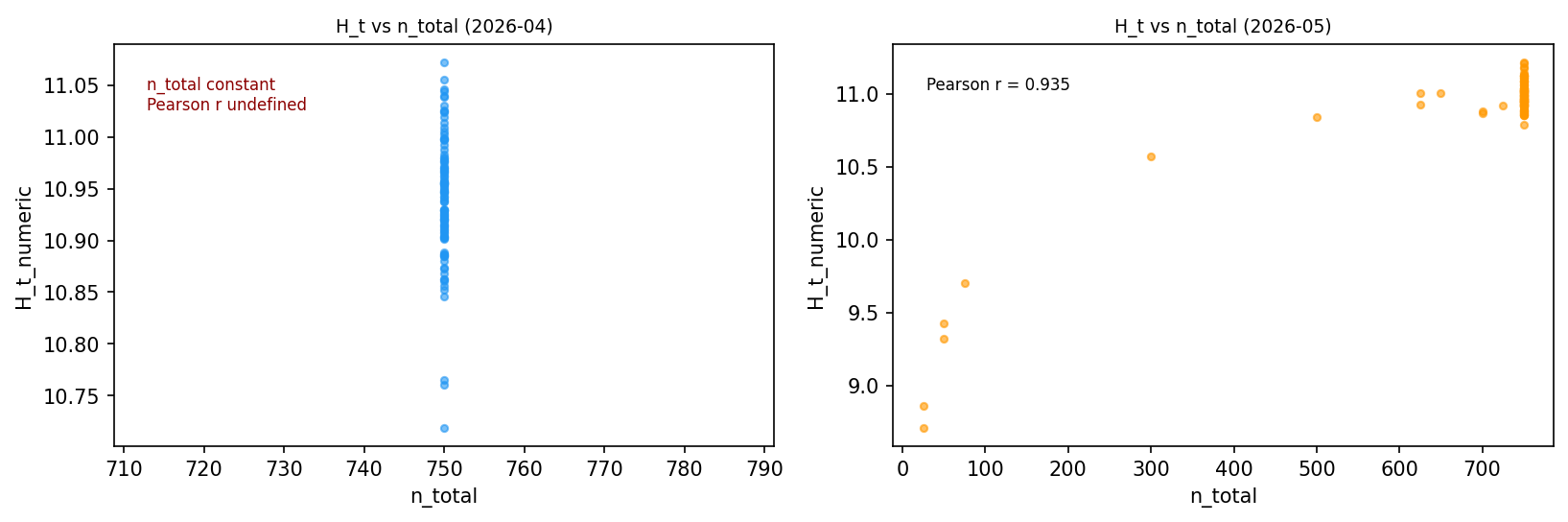}
\caption{$\Ht$ versus $n_{\mathrm{total}}$, separated by month. April has constant $n_{\mathrm{total}}$, so the April-only Pearson correlation is undefined; May shows the strong positive dependence reported in Table~\ref{tab:phase2-ht-ntotal-corr}.}
\label{fig:ht-ntotal}
\end{figure}

In April, $n_{\mathrm{total}}$ is constant at $750$ for all $104$ usable rows, and the Pearson correlation $r$ for April is undefined. In May, where $n_{\mathrm{total}}$ varies, raw $\Ht$ is strongly correlated with sample support ($r=0.9353$, $p=1.53\times 10^{-50}$). The two-month correlation is similarly high ($r=0.9182$, $p=3.10\times 10^{-87}$) and is driven by the May variation in $n_{\mathrm{total}}$. Figure~\ref{fig:ht-ntotal} shows this difference in sample support between the two months.

\begin{table}[h]
\centering
\small
\setlength{\tabcolsep}{4pt}
\caption{$\Ht$-$n_{\mathrm{total}}$ correlation.}
\label{tab:phase2-ht-ntotal-corr}
\begin{adjustbox}{max width=\textwidth}
\begin{tabular}{lllll}
\toprule
Scope & $n$ pairs & Pearson $r$ & $p$ & Note \\
\midrule
April-only & 104 & undefined & - & \makecell[l]{undefined ($n_{\mathrm{total}}$ constant at 750\\ for all 104 usable April rows)} \\
May-only & 110 & 0.9353 & 1.53e-50 & $n_{\mathrm{total}}$ varies in May \\
Two-month & 214 & 0.9182 & 3.10e-87 & driven by May variation \\
\bottomrule
\end{tabular}
\end{adjustbox}
\end{table}

These correlations set the interpretation boundary for raw $\Ht$. When sample support varies, raw $\Ht$ must be read together with $n_{\mathrm{total}}$. This dependence motivates the rarefied entropy measurement in Section~\ref{sec:rarefied-entropy}.

%===================================

\subsection{Rarefied lexical entropy}\label{sec:rarefied-entropy}

The dependence above makes a common token budget necessary for comparing lexical entropy across runs. Therefore, we recompute lexical entropy by rarefaction. 
%For each run, random subsamples of a fixed size are drawn from the stored token distribution, and Shannon entropy is computed for each subsample.

For each run, the stored top $200$ unigram counts define the probability of selecting each token type. We generate $50$ random samples of $B$ tokens from this distribution, with each token type selected according to its observed frequency. Each such sample is a multinomial draw. Shannon entropy is computed for every sample, and their mean entropy in bits gives the rarefied entropy $H_{\mathrm{rare}}$.

The pseudorandom number generator is initialized with seed $42$, so the same samples are reproduced when the calculation is repeated. The primary budget is $B=703$ tokens, equal to the smallest stored top $200$ token total among the $214$ valid rows. A secondary budget, $B_2=5000$ tokens, is used for the $209$ rows with at least $5000$ stored top $200$ tokens to test sensitivity to the chosen budget.

Because only the top $200$ unigram types are stored for each run, the rarefied entropy $H_{\mathrm{rare}}$ measures the evenness of that truncated distribution at a fixed token budget, and it differs from entropy computed on the full vocabulary. A lexical entropy metric that avoids the top $200$ truncation would require full unigram counts.

Table~\ref{tab:phase3-hrare-ntotal-corr} reports the sample support check for raw $\Ht$ and for rarefied entropy $H_{\mathrm{rare}}$ at the two rarefaction budgets. Pearson correlation $r$ measures linear association. Spearman rank correlation $\rho$ measures monotonic association through ranks \citep{spearman1904proof}. A rank is the position of an observation after the values are sorted from smallest to largest. The Spearman $p$ value indicates how surprising a correlation this far from zero would be if no monotonic association were present. Smaller $p$ values provide stronger evidence that the monotonic association is not zero.

\begin{table}[H]
\centering
\small
\setlength{\tabcolsep}{4pt}
\caption{Correlations between rarefied entropy $H_{\mathrm{rare}}$ and $n_{\mathrm{total}}$ by scope and rarefaction budget, with raw $\Ht$ rows for reference. Budget $B=703$ is the minimum stored top $200$ token total across the 214 valid rows. Budget $B_2=5000$ is used for the 209 rows with at least $5000$ stored top $200$ tokens and serves as a sensitivity check for the budget. The April-only row at $B=703$ is undefined because $n_{\mathrm{total}}=750$ for every usable April observation. The same applies to April-only correlations at other rarefaction budgets.}
\label{tab:phase3-hrare-ntotal-corr}
\begin{adjustbox}{max width=\textwidth}
\begin{tabular}{llrrrrr}
\toprule
Scope & Budget & $n$ & Pearson $r$ & Pearson $p$ & Spearman $\rho$ & Spearman $p$ \\
\midrule
April-only & $B=703$ & 104 & undefined & - & undefined & - \\
May-only & $B=703$ & 110 & $-0.4745$ & 1.63e-07 & 0.0124 & 0.898 \\
Two-month & $B=703$ & 214 & $-0.1768$ & 0.00955 & 0.1241 & 0.0699 \\
May-only & $B_2=5000$ & 105 & 0.3347 & 4.84e-04 & 0.3164 & 0.00101 \\
Two-month & $B_2=5000$ & 209 & 0.2837 & 3.14e-05 & 0.2813 & 3.69e-05 \\
\midrule
May-only & raw $\Ht$ & 110 & 0.9353 & 1.53e-50 & 0.4227 & 4.22e-06 \\
Two-month & raw $\Ht$ & 214 & 0.9182 & 3.10e-87 & 0.2747 & 4.62e-05 \\
\bottomrule
\end{tabular}
\end{adjustbox}
\end{table}

Raw $\Ht$ changes strongly with $n_{\mathrm{total}}$, indicating that runs with larger samples tend to appear more lexically diverse. Rarefaction compares samples at a common token budget and greatly weakens this dependence. A residual relation with sample support remains, and its form depends on the budget. At $B=703$, the two association measures disagree in May. The Pearson correlation is negative ($r=-0.4745$), while the Spearman rank correlation is near zero ($\rho=0.0124$), and no consistent monotonic ordering of $H_{\mathrm{rare}}$ with sample support is present. Excluding the six May observations with $n_{\mathrm{total}}\leq300$ reverses the sign of the Pearson coefficient ($r=0.3103$). Thus, the negative May Pearson association is concentrated in those six observations. In the restricted sample, $n_{\mathrm{total}}$ equals $750$ in $97$ of the $104$ observations. The variation in sample support is supplied by only seven observations, and the restricted Pearson coefficient carries little information about a general positive relation. At $B_2=5000$, both correlations are positive and similar in magnitude. 
%The residual relation with sample support depends on the rarefaction budget and on the range of support present in the observations. 
The budget and the range of sample support present in the observations remain part of the definition and interpretation of $H_{\mathrm{rare}}$.

\begin{figure}[h]
\centering
\includegraphics[width=0.94\textwidth]{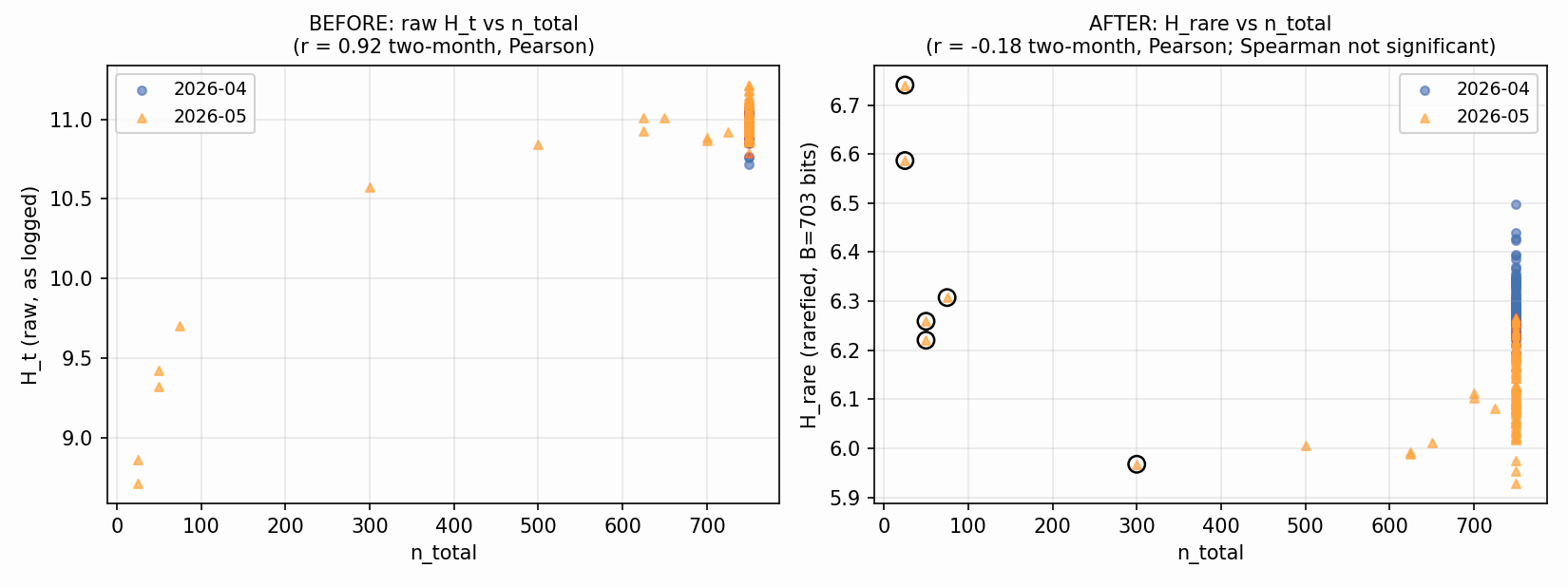}
\caption{Check of sample support before and after rarefaction. Rarefied entropy is computed from the stored top $200$ unigram distribution at $B=703$ tokens using $50$ random samples and seed $42$. The left panel shows raw $\Ht$ versus $n_{\mathrm{total}}$, and the right panel shows $H_{\mathrm{rare}}$ versus $n_{\mathrm{total}}$, separated by month. Circled points indicate observations with $n_{\mathrm{total}}\leq 300$.}
\label{fig:hrare-ntotal}
\end{figure}

\begin{figure}[htbp]
\centering
\includegraphics[width=0.92\textwidth]{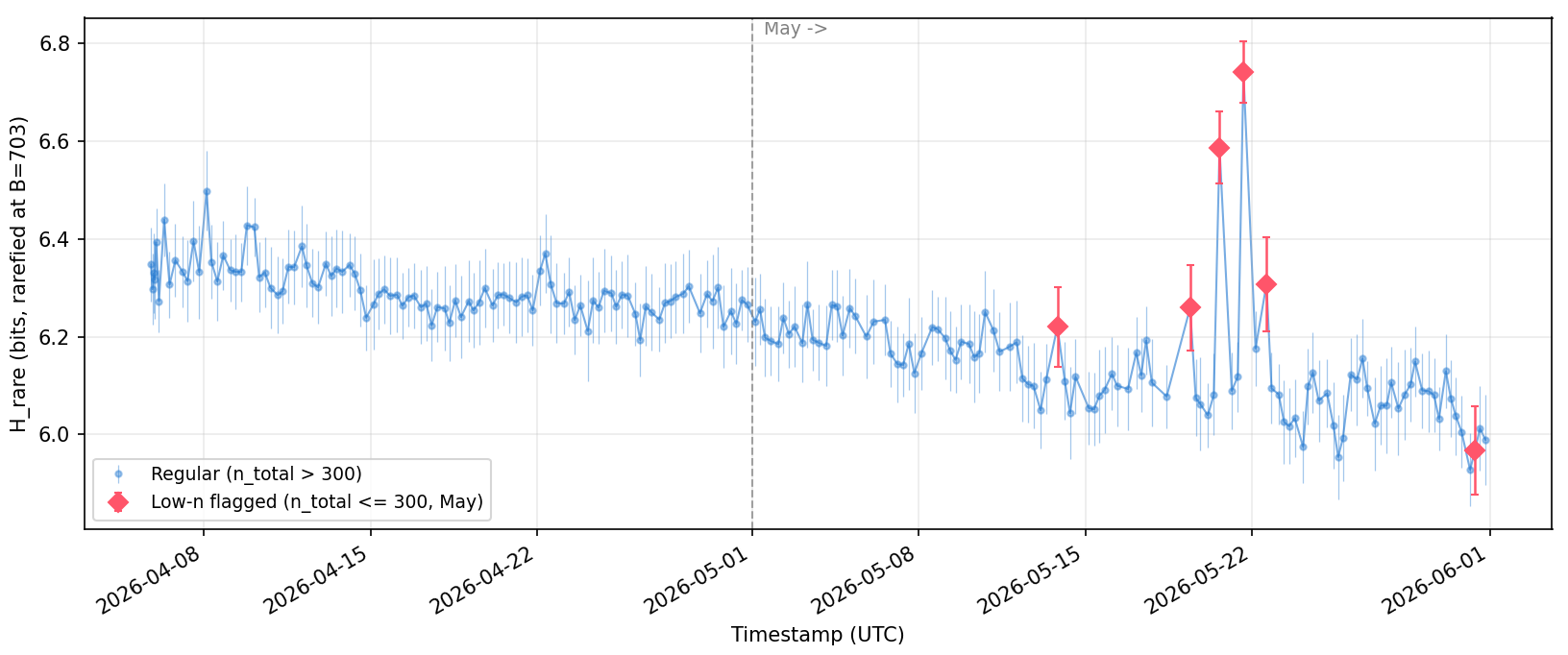}
\caption{Rarefied lexical entropy $H_{\mathrm{rare}}$ computed from the stored top $200$ unigram distribution at $B=703$ tokens using $50$ random samples and seed $42$ over the observation window. Marked points correspond to rows with $n_{\mathrm{total}}\leq 300$.}
\label{fig:hrare-timeseries}
\end{figure}

Figure~\ref{fig:hrare-ntotal} shows the relation with sample support before and after rarefaction. Figure~\ref{fig:hrare-timeseries} shows the rarefied entropy series over the observation window.

Rarefaction sets the reading rule for lexical entropy when sample support varies. Raw $\Ht$ remains a useful observable when $n_{\mathrm{total}}$ is fixed or when the same change is also present in $H_{\mathrm{rare}}$ at a common token budget. Otherwise, lexical concentration is read through $H_{\mathrm{rare}}$.

%===================================

\subsection{April-May comparison at a common token budget}

To compare April and May at a common token budget, we use $H_{\mathrm{rare}}$ at the primary budget $B=703$ tokens. The monthly distributions are compared with the Mann-Whitney U test \citep{mann1947test}, which tests whether observations from one month tend to rank above or below observations from the other. The resulting $p$ value gives the probability of observing a separation between the April and May distributions at least this strong if no systematic difference were present. It is then adjusted with the Benjamini-Hochberg correction \citep{benjamini1995controlling}, applied jointly to all scalar measurements compared in April and May. The adjusted value is reported as $q$. 

Table~\ref{tab:phase3-hrare-april-may} reports the monthly means, the April-May shift, the shift relative to the two-month standard deviation, and the corresponding $p$ and $q$ values. 

\begin{table}[H]
\centering
\small
\setlength{\tabcolsep}{5pt}
\caption{April-vs-May comparison of rarefied lexical entropy at the common token budget $B=703$. The reported $q$ value retains the Benjamini-Hochberg correction applied jointly to all scalar measurements compared in April and May, including rarefied entropy.}
\label{tab:phase3-hrare-april-may}
\begin{adjustbox}{max width=\textwidth}
\begin{tabular}{lcccccc}
\toprule
Metric & Mean April & Mean May & Shift & Shift/two-month SD & MW $p$ & BH $q$ \\
\midrule
$H_{\mathrm{rare}}$ & 6.2973 & 6.1373 & $-0.1600$ & $-1.374$ & $4.36\times10^{-31}$ & $3.92\times10^{-30}$ \\
\bottomrule
\end{tabular}
\end{adjustbox}
\end{table}

The mean $H_{\mathrm{rare}}$ decreases from $6.297$ bits in April to $6.137$ bits in May, giving an April-May shift of $-0.160$ bits. This shift is $-1.374$ times the two-month standard deviation, so it is large relative to the observed run-to-run variation. The Mann-Whitney test gives $p=4.4\times 10^{-31}$, with Benjamini-Hochberg adjusted $q=3.9\times 10^{-30}$.

At the common token budget, the stored top $200$ token distribution is less even in May than in April. More of the observed token mass is concentrated in fewer token types. The very small $p$ and $q$ values indicate that the separation between the two monthly distributions is far larger than expected if there were no systematic April-May difference. This supports a measurable increase in lexical concentration in May, within the stored top $200$ distribution.

%===================================

\subsection{Sensitivity to sparse observations}

Six May observations have $n_{\mathrm{total}}\leq 300$. To measure their influence, we repeat the April-May comparison after excluding them. At budget $B=703$, the May mean $H_{\mathrm{rare}}$ changes from $6.1373$ to $6.1252$ bits. The April-May shift changes from $-0.1600$ to $-0.1721$ bits. The Mann-Whitney probability changes from $4.36\times 10^{-31}$ to $1.66\times 10^{-33}$. This means that the decrease in rarefied entropy remains and becomes slightly larger after the six observations are excluded.

Raw $\Ht$ responds differently. Its May mean changes from $10.9056$ to $10.9906$ bits, and the April-May shift reverses from $-0.0349$ to $0.0501$ bits. This reversal is consistent with the dependence of raw $\Ht$ on sample support and shows that the raw April-May comparison is not stable under the exclusion. Table~\ref{tab:lexical-low-n-sensitivity} reports the monthly comparison for both quantities.

\begin{table}[h]
\centering
\small
\setlength{\tabcolsep}{4pt}
\caption{Effect of excluding the six May low-$n_{\mathrm{total}}$ observations on raw and rarefied lexical entropy. April contains $104$ observations, May contains $110$ observations, and the reduced May sample contains $104$ observations. Rarefied entropy is computed at budget $B=703$.}
\label{tab:lexical-low-n-sensitivity}
\begin{adjustbox}{max width=\textwidth}
\begin{tabular}{lrrrrr}
\toprule
Metric
& April mean
& May mean
& May mean excl.
& April-May shift
& Shift after excl. \\
\midrule
$H_{\mathrm{rare}}$ ($B=703$)
& 6.2973
& 6.1373
& 6.1252
& $-0.1600$
& $-0.1721$ \\
$\Ht$
& 10.9405
& 10.9056
& 10.9906
& $-0.0349$
& 0.0501 \\
\bottomrule
\end{tabular}
\end{adjustbox}
\end{table}

Across the full two-month sample, the exclusion changes the mean $H_{\mathrm{rare}}$ by only $-0.0038$ bits, which is $3.3\%$ of the two-month standard deviation. The mean raw $\Ht$ changes by $0.0430$ bits, which is $15.2\%$ of the two-month standard deviation. The six observations remain included in the primary analysis.

%===================================

\subsection{Sensitivity to mint payload tokens}
\label{sec:mint-payload-sensitivity}

The sampled feed contains mechanically repeated JSON payloads associated with the mint operations for \texttt{claw}, \texttt{mbc20}, and \texttt{k0rp}. To test whether these payloads account for the April-May entropy difference, we subtract the identified compact JSON token strings for these mint operations from the stored top $200$ unigram counts and recompute $H_{\mathrm{rare}}$ with the same rarefaction procedure at budget $B=703$.

The filter removes $967$ token occurrences, corresponding to $0.055\%$ of the stored token mass. Mint payloads form the largest class of such repeated payload tokens in the stored counts and are used here to test the effect of the dominant payload contribution. The audit also identifies $198$ occurrences produced by transfer payloads. These are reported separately and remain in the counts.

The form of the payload tokens differs between the two months. In April, compact JSON payloads remain whole under the whitespace tokenizer. In May, spaces within the JSON separate the payloads into field fragments, six of which appear in the stored top $200$ counts. While the payload formatting changes between April and May, the tokenizer remains unchanged. The analysis does not identify the cause of this formatting change.

Table~\ref{tab:mint-payload-sensitivity} reports the change in mean $H_{\mathrm{rare}}$ after the identified mint payload tokens are subtracted. The subtraction lowers rarefied entropy in every comparison window. The mean change is $-0.001475$ bits in April and $-0.004949$ bits in May. Thus, the identified mint payload tokens slightly increase entropy within the stored top $200$ distributions.

\begin{table}[h]
\centering
\small
\setlength{\tabcolsep}{4pt}
\caption{Sensitivity of rarefied entropy to the identified mint payload tokens. The reported change is the mean $H_{\mathrm{rare}}$ after subtraction minus the mean before subtraction. Negative values indicate that subtracting the payload tokens lowers entropy. Rarefied entropy is computed at budget $B=703$.}
\label{tab:mint-payload-sensitivity}
\begin{adjustbox}{max width=\textwidth}
\begin{tabular}{lr}
\toprule
Scope & Mean change after subtraction (bits) \\
\midrule
April-only & $-0.001475$ \\
May-only & $-0.004949$ \\
Two-month & $-0.003261$ \\
Two-month excluding low-$n_{\mathrm{total}}$ observations & $-0.002661$ \\
\bottomrule
\end{tabular}
\end{adjustbox}
\end{table}

The subtraction lowers the May mean by $0.003474$ bits more than the April mean. The April-May decrease in $H_{\mathrm{rare}}$ therefore becomes slightly larger after the mint payload tokens are removed. This differential effect is about $2.2\%$ of the observed $0.1600$ bit April-May shift.

The identified mint payload tokens do not account for the measured increase in lexical concentration in May. Their measured effect has the opposite direction and is small relative to the cross-month difference. The effect also remains small when the six low-$n_{\mathrm{total}}$ observations are excluded.

%%%%%%%%%%%%%%%%%%%%%%%%%%%%%%%%%%%%%%%
%%%%%%%%%%%%%%%%%%%%%%%%%%%%%%%%%%%%%%%

\section{Cross-layer geometry}\label{sec:embedding-geometry}

Dividing each feed observation by content age produces three simultaneous semantic layers whose relative positions can be measured within the same run. Each stored text is represented in a common $384$ dimensional sentence embedding space, where vectors corresponding to semantically similar texts tend to point in similar directions \citep{reimers2019sentencebert}. The surface, mid-stream, and residue centroids provide a geometric description of how recent and older content are organized around the semantic state of each run.

We first measure how each layer moves relative to the run mean. Then, we determine which features of the residual geometry arise from the centering construction itself. Finally, deterministic shuffling and cross-month replication test whether recurring distance relations among the layers persist beyond a local observation period. These analyses show what the age partition makes measurable while setting the boundary between resolved geometric organization and unsupported readings.

%===================================

\subsection{Motion relative to the run mean}\label{sec:layer-resid-geom}

The run mean describes the shared semantic position of the sampled feed within each observation. Subtracting this common center from each layer centroid gives the layer residual $r_\ell(t)$ defined in Eq.~\eqref{eq:layer-resid}. Its norm measures how far the layer lies from the shared state of the run, while changes in the residual vector across adjacent runs measure how the layer moves relative to that state as the feed changes over time.

The residual sum relation in Eq.~\eqref{eq:layer-residual-sum} shows that the residuals form a zero sum configuration when the run mean embedding $\bar e(t)$ equals the equal weight centroid mean $\bar c(t)$. Across the $210$ rows with all three layer centroids valid, the median value of norm $\|\bar e(t)-\bar c(t)\|_2$ is $1.39\times 10^{-7}$, and $203$ rows have a gap no larger than $10^{-6}$. The remaining seven rows, all in May, have gaps from $1.85\times 10^{-3}$ to $7.50\times 10^{-2}$.

Thus, the three residuals form an approximately zero sum configuration in most observations. This condition constrains their joint geometry and becomes important when the residual angles are interpreted below.

\begin{figure}[H]
\centering
\includegraphics[width=1\textwidth]{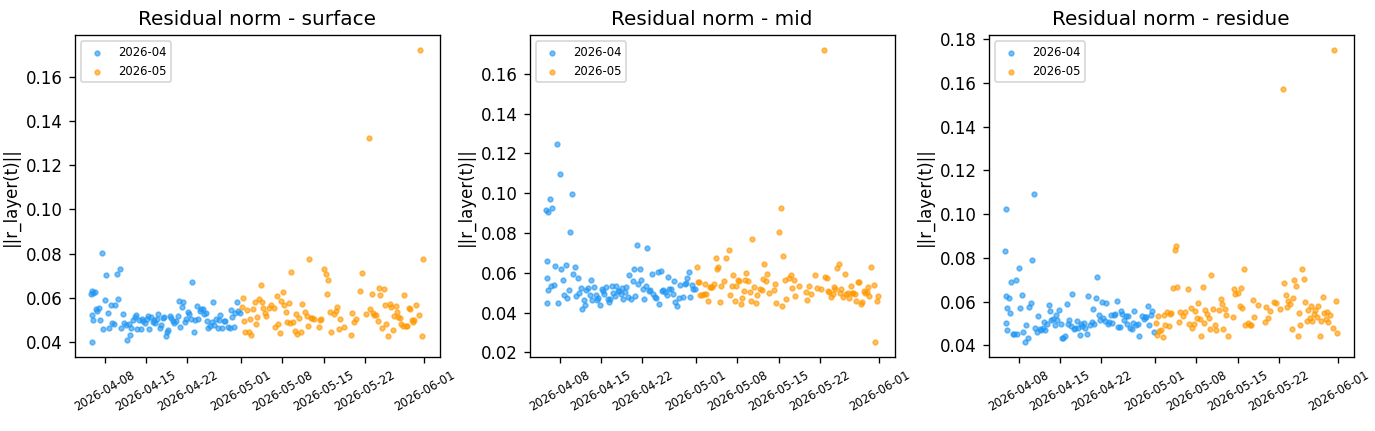}
\caption{Distributions of layer residual norms across the $210$ observations with valid surface, mid-stream, and residue centroids.}
\label{fig:residual-geometry-norms}
\end{figure}

Across the two-month record, the three layers lie at similar average distances from the run mean. Their mean residual norms are $0.0537$ for the surface layer, $0.0553$ for the mid-stream layer, and $0.0561$ for the residue layer. The corresponding distributions are shown in Figure~\ref{fig:residual-geometry-norms}.

\begin{figure}[H]
\centering
\includegraphics[width=1\textwidth]{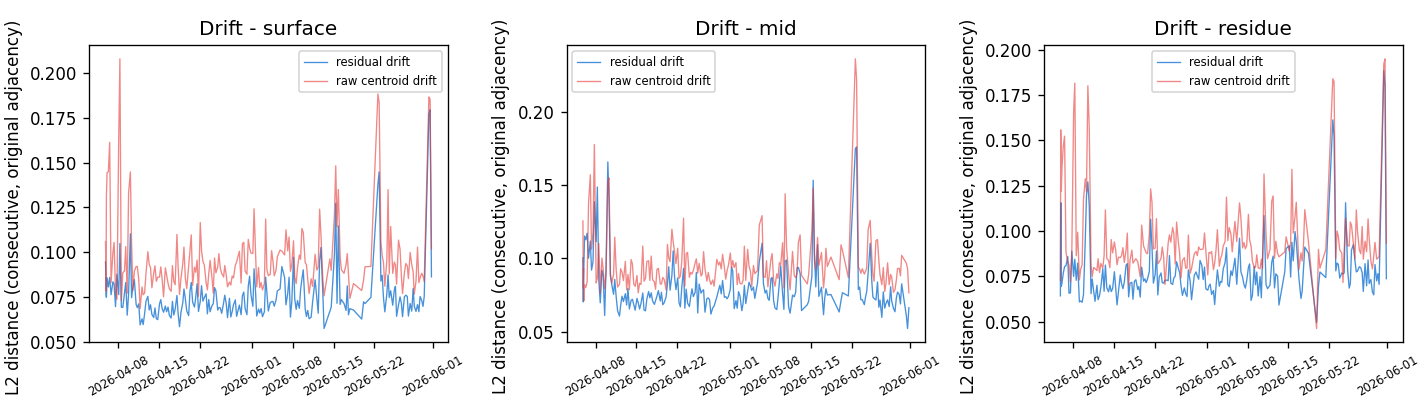}
\caption{Residual drift compared with raw layer centroid drift.}
\label{fig:residual-geometry-drift}
\end{figure}

Relative layer motion is measured over the $204$ adjacent pairs in the original observation sequence for which both endpoints have valid layer centroids. Missing centroid rows are not bridged. The overall count includes one April-May boundary pair, which is not assigned to either monthly row. Writing each pair as $t_j$ and $t_{j+1}$, the residual drift of layer $\ell$ is
\begin{equation}
\Delta r_\ell(t_j)
=
\left\|
r_\ell(t_{j+1})-r_\ell(t_j)
\right\|_2,
\end{equation}
and the corresponding raw centroid drift is
\begin{equation}
\Delta c_\ell(t_j)
=
\left\|
c_\ell(t_{j+1})-c_\ell(t_j)
\right\|_2.
\end{equation}
Their ratio,
\begin{equation}
R_\ell(t_j)
=
\frac{\Delta r_\ell(t_j)}
{\Delta c_\ell(t_j)},
\end{equation}
compares the scale of motion relative to the run mean with the full displacement of the layer centroid. Table~\ref{tab:phase2-residual-geometry-drift} summarizes the residual drifts and their ratios to raw centroid drift. Figure~\ref{fig:residual-geometry-drift} shows residual drift against raw centroid drift across adjacent pairs.

\begin{table}[H]
\centering
\small
\setlength{\tabcolsep}{5pt}
\renewcommand{\arraystretch}{1.08}
\caption{Residual motion relative to raw layer centroid motion over adjacent valid observations.}
\label{tab:phase2-residual-geometry-drift}
\begin{adjustbox}{max width=\textwidth}
\begin{tabular}{@{}lccccccc@{}}
\toprule
Scope &
$n$ &
\makecell{Surface\\residual drift} &
\makecell{Surface\\ratio} &
\makecell{Mid-stream\\residual drift} &
\makecell{Mid-stream\\ratio} &
\makecell{Residue\\residual drift} &
\makecell{Residue\\ratio} \\
\midrule
overall & 204 & 0.0757 & 0.793 & 0.0802 & 0.836 & 0.0777 & 0.822 \\
April   & 103 & 0.0729 & 0.789 & 0.0809 & 0.857 & 0.0752 & 0.810 \\
May     & 100 & 0.0786 & 0.799 & 0.0796 & 0.816 & 0.0804 & 0.833 \\
\bottomrule
\end{tabular}
\end{adjustbox}
\par\vspace{0.35em}
\parbox{\textwidth}{\footnotesize\emph{Note.}
Residual drift and raw centroid drift are Euclidean displacements between adjacent valid observations. The ratio is residual drift divided by raw centroid drift. Missing centroid rows are not bridged.}
\end{table}

Across the same $204$ adjacent pairs, the mean displacement of the run mean is $0.0568$. The mean residual/raw ratios range from $0.79$ to $0.84$ across layers in the full record and remain similar within April and May. Therefore, residual motion remains comparable in scale to the full centroid displacement after movement of the run mean is removed. The three post age layers move with the shared semantic state of the sampled feed and also change substantially relative to it. This relative motion provides the geometric quantity needed to test propagation or subgroup formation once directed transitions or persistent membership information are measured alongside. It also supports testing influence when combined with a controlled intervention.

The pairwise residual angles provide a check on the centering construction. Their mean across the valid layer pairs is $2.094$ rad, or $120.0^\circ$. A Gaussian reference formed from independent random vectors rescaled to the observed residual norms has a mean angle of $1.571$ rad, close to $90^\circ$. Independent directions in the $384$ dimensional embedding space concentrate near $90^\circ$, while the residual sum relation in Eq.~\eqref{eq:layer-residual-sum} constrains the three centered layer vectors to a different geometry.

When three vectors have similar norms and form an approximately zero sum configuration, their pairwise inner products approach one half of the product of their norms with opposite sign. Their angles then approach
\begin{equation}
\arccos\left(-\frac{1}{2}\right)
=
\frac{2\pi}{3}
=
120^\circ.
\end{equation}
A more detailed derivation of both geometric reference angles is given in Appendix~\ref{app:residual-angle-geometry}.

The observed mean angle is consistent with the algebraic constraint created by centering three layer centroids around a common mean. The $120^\circ$ relation serves as a baseline of the residual construction. Departures from this baseline become measurable when differences in residual norms and the gap between $\bar e(t)$ and $\bar c(t)$ are included in the comparison.

Together, these results distinguish measured relative motion from geometry imposed by centering. The residual displacements show that the post age layers change relative to the shared state of each run. The angles near $120^\circ$ define the baseline created by the residual construction. The controls below ask whether pairwise layer separation contains a recurring relation associated with post age beyond this baseline.

%===================================

\subsection{Layer separation under deterministic shuffling}

The deterministic shuffle provides an internal reference for the cross-layer distances produced within the same observations. Comparing the observed and shuffled distances tests whether separation among the surface, mid-stream, and residue centroids exceeds the scale produced by the shuffle procedure. A recurring excess would support geometric organization associated with post age at the resolution of this control.

The observation schedule contains $14$ shuffle checks. One April check lacks usable centroid distance comparisons, leaving $13$ usable entries, with $8$ in April and $5$ in May. Each usable entry gives observed and shuffled values for $d_{SM}$, $d_{MR}$, and $d_{SR}$, producing $39$ paired distance comparisons.
The clearest departure from the shuffle reference occurs for the mid-residue pair. The observed distance $d_{MR}$ exceeds its shuffled value in $10$ of the $13$ entries, with a mean observed-minus-shuffled difference of $0.0166$. The corresponding counts are $6$ of $13$ for $d_{SM}$ and $7$ of $13$ for $d_{SR}$, with median differences close to zero. 

Across the $13$ available shuffle entries, the control resolves a recurring excess in mid-residue separation. The same directional relation is not resolved for $d_{SM}$ or $d_{SR}$, which prevents a reading of stable geometric organization across all three post age layers. Continued scheduled shuffle checks provide a direct test of whether the mid-residue excess persists and whether recurring relations become measurable for the other two layer pairs.

\subsection{Cross-month stability of layer ordering}
\label{sec:layer-ordering-crossrun}

The three layers, labeled $S$, $M$, $R$, define three pairwise distances, $d_{SM}$, $d_{MR}$, and $d_{SR}$. Ranking these three distances from smallest to largest gives $3!=6$ possible orderings. For example, one possible ordering is
\begin{equation*}
d_{SM}<d_{MR}<d_{SR}.
\end{equation*}

A global label permutation check compares the observed ordering statistics with the values obtained under all six possible relabelings of $S$, $M$, and $R$, including the original labeling. The most frequent ordering in April is then carried forward as a candidate pattern and tested for recurrence in May.
The minimum achievable two-sided tail probability of the global permutation test with only six null labelings is
\begin{equation*}
p_{\min}
=
\frac{1}{6}
=
0.1667.
\end{equation*}
Thus, the test measures whether the original labeling is unusual within this exact six-point null set. Note that it cannot reach the conventional threshold of $0.05$.

\begin{table}[htbp]
\centering
\small
\setlength{\tabcolsep}{4pt}
\caption{Global permutation check for layer distance ordering across the $210$ observations with valid layer centroids.}
\label{tab:phase2-layer-order-permutation}
\begin{adjustbox}{max width=\textwidth}
\begin{tabular}{@{}lllllll@{}}
\toprule
Statistic & Observed &
\makecell[l]{Exact null\\mean} &
\makecell[l]{Exact null\\std.} &
\makecell[l]{Two-sided\\\,tail $p$} &
\makecell[l]{$n$ null\\perms.} &
\makecell[l]{$n$\\used} \\
\midrule
$P(d_{SM} < d_{MR})$            & 0.609524 & 0.5      & 0.080906 & 0.3333   & 6 & 210 \\
 $P(d_{SR}\ \text{is largest})$ & 0.309524 & 0.333333 & 0.051434 & 1        & 6 & 210 \\
$P(d_{SM} < d_{MR} < d_{SR})$   & 0.204762 & 0.166667 & 0.045675 & 0.6667   & 6 & 210 \\
\bottomrule
\end{tabular}
\end{adjustbox}
\end{table}
Table~\ref{tab:phase2-layer-order-permutation} compares the observed ordering frequencies with the values obtained under the six possible assignments of the layer labels. Across the $210$ valid observations, $d_{SM}<d_{MR}$ occurs in $128$ runs, or $61.0\%$; $d_{SR}$ is the largest distance in $65$ runs, or $31.0\%$; and the full ordering $d_{SM}<d_{MR}<d_{SR}$ occurs in $43$ runs, or $20.5\%$. The corresponding expectations across label permutations are $1/2$, $1/3$, and $1/6$.

The exact tail probabilities are $0.3333$, $1$, and $0.6667$. The original assignment of the surface, mid-stream, and residue labels is not distinguished from the other five assignments by these ordering statistics. In addition, as noted above, the resolution of the test is limited by its six possible permutations, which set the minimum two-sided tail probability at $0.1667$.

The permutation check does not support an association between post age labels and a particular distance ordering across the full record. To test the inter-layer geometric patterns further, we select an ordering in one observation period and check whether it recurs in a separate period. In particular, we identify the most frequent ordering in April and compare its frequency in May with the uniform expectation of $1/6$ across the six possible orderings. Recurrence above this expectation would support persistence of the geometric relation across the two observation periods.

The most frequent April ordering is
\begin{equation*}
d_{SR}<d_{MR}<d_{SM},
\end{equation*}
with surface and residue closest and surface and mid-stream farthest. It occurs in $25$ of $104$ April observations, or $24.0\%$. In May, the same ordering occurs in $11$ of $106$ observations, or $10.4\%$, below the uniform expectation of $1/6=16.7\%$. The one-sided exact binomial test gives $p=0.975$ and does not support recurrence above the uniform expectation. The most frequent May ordering is instead
\begin{equation*}
d_{SM}<d_{SR}<d_{MR}.
\end{equation*}

A secondary omnibus $\chi^2$ goodness-of-fit test compares the frequencies of all six May orderings with a uniform distribution. When all $106$ May observations are treated as independent, the test gives $\chi^2(5)=16.49$ and $p=0.0056$, which supports a nonuniform ordering distribution under that assumption. Successive observations are autocorrelated because posts can remain in the sampled feed, move between age layers, and contribute to persistent feed composition. The effective number of independent observations is smaller than $106$, and the result from the full sample may overstate the evidence for nonuniformity.

To reduce this dependence, we repeat both May tests after retaining every fourth observation, which gives approximately one observation per day. The most frequent April ordering occurs in $2$ of the $27$ retained observations, and the exact binomial test gives $p=0.953$. The omnibus test gives $\chi^2(5)=9.67$ and $p=0.0853$. Each expected cell count is only $27/6=4.5$. The $\chi^2$ approximation is coarse at this sample size. The thinned results support neither recurrence of the April ordering nor a clear departure of the May ordering distribution from uniform. Table~\ref{tab:phase3-layer-ordering} reports the full comparison.

\begin{table}[h]
\centering
\small
\setlength{\tabcolsep}{4pt}
\caption{Cross-run layer ordering consistency. The modal inter-layer distance ordering is identified on April rows and tested on May rows. The primary test is a one-sided exact binomial against the null proportion $1/6$; the secondary test is a $\chi^2$ goodness-of-fit of the full May distribution against uniform, with a thinned version that keeps every fourth run ($\approx$ 1 run/day) as a coarse check for autocorrelation.}
\label{tab:phase3-layer-ordering}
\begin{adjustbox}{max width=\textwidth}
\begin{tabular}{lllll}
\toprule
Test & Ordering / scope & Observed & Statistic & $p$ \\
\midrule
April modal ordering & $d_{SR}<d_{MR}<d_{SM}$ & 25/104 (24.0\%) & - & - \\
May occurrence of April modal & $d_{SR}<d_{MR}<d_{SM}$ & 11/106 (10.4\%) & exact binomial  & 0.975 \\
May omnibus vs uniform & all six orderings & 106 runs & $\chi^2(5)=16.49$ & 0.00557 \\
May omnibus (every 4th run) & all six orderings & 27 runs & $\chi^2(5)=9.67$ & 0.0853 \\
May occurrence (every 4th run) & $d_{SR}<d_{MR}<d_{SM}$ & 2/27 (7.4\%) & exact binomial  & 0.953 \\
\bottomrule
\end{tabular}
\end{adjustbox}
\end{table}

The April pattern did not carry into May. In both the full May sequence and the thinned sequence, it occurred less often than the uniform expectation of $1/6$. The present observations do not support a stable layer ordering across periods. Such a claim becomes possible only if the same ordering recurs across additional observation periods under checks for temporal dependence. At the current level of support, the instrument already applies the cross-period test that a stable layer ordering would have to pass.

%%%%%%%%%%%%%%%%%%%%%%%%%%%%%%%%%%%%%%%
%%%%%%%%%%%%%%%%%%%%%%%%%%%%%%%%%%%%%%%

\section{Temporal organization of the measured quantities}
\label{sec:lexical-temporal}

In this section, we address two temporal questions. First, are consecutive state rows lexically closer than rows made adjacent after timestamp shuffling? Second, do detrended semantic homogeneity and run-to-run drift support division into intervals with different local statistical behavior? The first analysis tests temporal dependence in the token distribution. The second locates candidate boundaries in the measured series. Together, they move the instrument from measuring change between observations to locating where the local behavior of that change may shift.

%===================================

\subsection{Lexical temporal dependence}

Lexical change between consecutive state rows is measured from their stored top $200$ unigram count vectors. Jensen-Shannon (JS) distance compares how the relative token frequencies are distributed across two runs. It is zero when the two runs assign the same fraction of their stored token mass to every aligned token type, and it increases as those allocations differ. The comparison is symmetric, so exchanging the order of the two runs does not change the distance \citep{lin1991divergence}.

Let $u(t)$ and $u(t^-)$ denote the unigram count vectors for run $t$ and the preceding run, with their components aligned by token. For each token $w$, define the normalized frequency
\begin{equation}
\widetilde p_t(w)
=
\frac{u_w(t)}{\sum_v u_v(t)}.
\end{equation}
The collection of these frequencies forms the truncated empirical token distribution for run $t$. The midpoint distribution assigns each token the average of its frequencies in the two runs,
\begin{equation}
m_t(w)
=
\frac{1}{2}
\left[
\widetilde p_t(w)+\widetilde p_{t^-}(w)
\right].
\end{equation}
The Jensen-Shannon distance is
\begin{equation}
d_{\mathrm{JS}}(t)
=
\left[
\frac{1}{2}
\sum_w
\widetilde p_t(w)
\log_2
\frac{\widetilde p_t(w)}{m_t(w)}
+
\frac{1}{2}
\sum_w
\widetilde p_{t^-}(w)
\log_2
\frac{\widetilde p_{t^-}(w)}{m_t(w)}
\right]^{1/2}.
\end{equation}
The two terms measure how far each run lies from their shared midpoint distribution. Taking their average and square root produces a bounded distance between the two lexical states. The consecutive lexical cosine step $d_{\mathrm{lex}}(t)$, defined in Section~\ref{sec:measured-quantities}, compares the corresponding aligned unigram count vectors.

To construct the reference, we permute timestamp order $200$ times while keeping each stored unigram count vector intact. For each permutation, we recompute JS and cosine distances between rows made adjacent by the shuffled order. The observed consecutive distances are then compared with the reference distribution of per-shuffle mean distances. Lower observed values indicate that true temporal neighbors have more similar lexical states than pairings produced after temporal order is destroyed. This comparison makes temporal dependence in token allocation measurable while leaving the process that produces it open.

\begin{table}[H]
\centering
\small
\setlength{\tabcolsep}{4pt}
\caption{Lexical temporal dependence relative to the timestamp shuffle
reference. Observed means and standard deviations are computed across
consecutive state row pairs. Each shuffled value is the mean distance across
the adjacent pairs produced by one timestamp permutation, and the shuffled
standard deviations are computed across the $200$ permutation means. Ratios
below $1$ indicate that true temporal neighbors are lexically closer than the
shuffled pairings.}
\label{tab:phase2-lexical-drift}
\begin{adjustbox}{max width=\textwidth}
\begin{tabular}{@{}lllllllllllll@{}}
\toprule
Scope &
\makecell[l]{$n$\\pairs} &
\makecell[l]{Observed\\JS mean} &
\makecell[l]{Observed\\JS std.} &
\makecell[l]{Observed\\cosine mean} &
\makecell[l]{Observed\\cosine std.} &
\makecell[l]{Shuffled\\JS mean} &
\makecell[l]{Shuffled\\JS std.} &
\makecell[l]{Shuffled\\cosine \\mean} &
\makecell[l]{Shuffled\\cosine \\std.} &
\makecell[l]{JS \\observed/\\shuffled} &
\makecell[l]{Cosine \\observed/\\shuffled} &
\makecell[l]{$n$\\shuffles} \\
\midrule
overall & 214 & 0.251273 & 0.0606756 & 0.00815225 & 0.00962067 & 0.318448 & 0.00173409 & 0.0161116 & 0.000306526 & 0.789056 & 0.505985 & 200 \\
2026-04 & 103 & 0.243634 & 0.0371922 & 0.00764776 & 0.00519231 & 0.318448 & 0.00173409 & 0.0161116 & 0.000306526 & 0.765067 & 0.474673 & 200 \\
2026-05 & 110 & 0.258721 & 0.0758116 & 0.00865751 & 0.0124181 & 0.318448 & 0.00173409 & 0.0161116 & 0.000306526 & 0.812441 & 0.537345 & 200 \\
\bottomrule
\end{tabular}
\end{adjustbox}
\end{table}

\begin{figure}[H]
\centering
\includegraphics[width=0.95\textwidth]{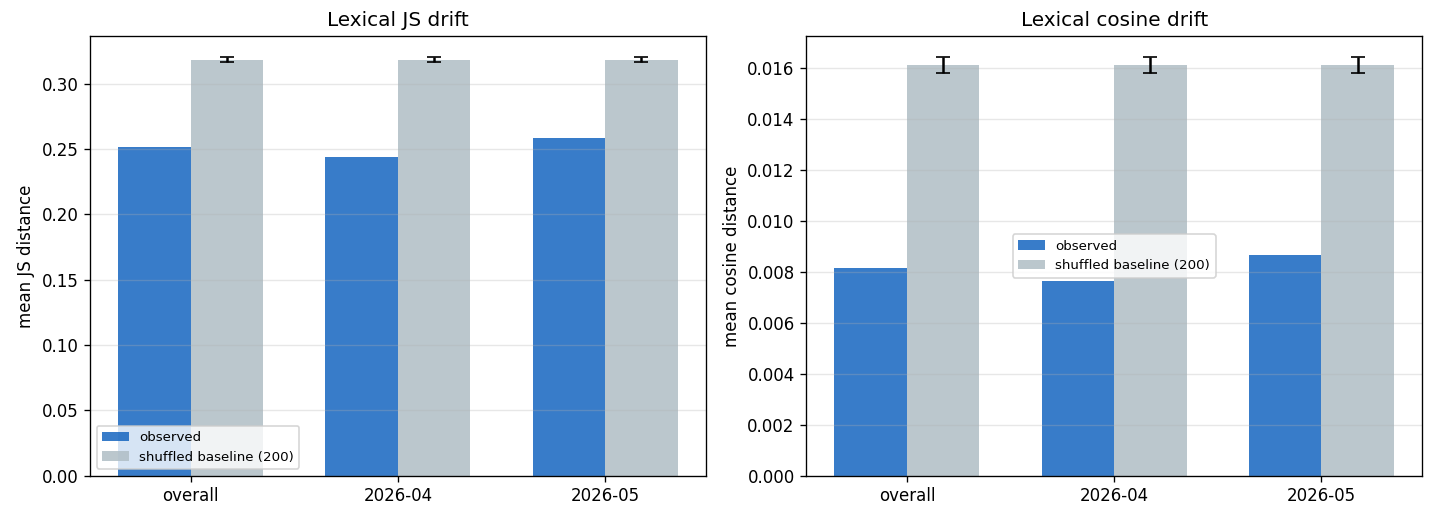}
\caption{Lexical temporal dependence relative to the timestamp shuffle
reference. Observed means below the shuffled means indicate that true temporal
neighbors have more similar lexical states than rows made adjacent after
timestamp permutation.}
\label{fig:lexical-drift}
\end{figure}

Table~\ref{tab:phase2-lexical-drift} reports the observed and shuffled mean distances for the full observation window, together with the April-only and May-only observed means. The monthly observed means use only consecutive pairs that remain within the same month. The April-May boundary pair is included only in the overall row. Both monthly rows are compared with the single shuffle reference constructed from the full two-month sequence. Figure~\ref{fig:lexical-drift} shows the same comparison as a bar chart.

Across the $214$ consecutive state row pairs in the full two-month sample,
the observed mean JS distance was $0.2513$, compared with $0.3184$
under timestamp permutation. The ratio of the observed mean to the shuffled
mean was $0.789$. The observed mean cosine distance was $0.00815$, compared
with a shuffled mean of $0.01611$, giving a ratio of $0.506$. Both measures
place true temporal neighbors closer than pairings produced after temporal
order is destroyed.

The April-only and May-only rows show the same ordering. For JS
distance, the observed-to-shuffled ratios were $0.765$ in April and $0.812$ in
May. For cosine distance, the corresponding ratios were $0.475$ and $0.537$.
Thus, the lexical proximity of true temporal neighbors is present in both
months under the common two-month shuffle reference.

The timestamp shuffle comparison supports temporal dependence in the stored
token distribution across the full two-month sample. It does not identify the
process that produces this dependence. We next ask whether semantic
homogeneity and run-to-run drift contain candidate boundaries between
intervals with different local statistical behavior.

%===================================

\subsection{Candidate segmentation}
\label{sec:changepoints}

After removing a fitted linear trend from semantic homogeneity $\phit$ and run-to-run drift $\Dt$ (detrending), we use changepoint analysis to locate positions where the local statistical behavior of either series changes. These positions divide the observation record into candidate intervals with different local behavior. Each boundary is a candidate segmentation point. Whether such a boundary corresponds to a transition in the collective semantic state is a further measurable question.

We apply the pruned exact linear time algorithm, PELT, separately to the detrended $\phit$ and $\Dt$ series \citep{killick2012optimal}. PELT searches for a set of boundaries that makes the observations within each resulting interval more statistically similar, while adding a penalty for every additional boundary. We use \texttt{ruptures} with \texttt{model="rbf"}, whose radial basis function cost responds to changes in the distribution of the series, and \texttt{min\_size=3}, which requires each interval to contain at least three runs \citep{truong2020selective}.

We select the penalty by sweeping candidate values and choosing the one with the minimum score modeled on the Bayesian information criterion (BIC), which balances segmentation fit against the number of resulting intervals \citep{schwarz1978estimating}. Sensitivity is evaluated at the selected penalty, that penalty divided by $\sqrt{3}$, and that penalty multiplied by $\sqrt{3}$. The procedure locates candidate boundaries in each detrended series.

At the selected penalty, the detrended $\phit$ series contains $12$
changepoints. The count changes from $22$ to $12$ to $5$ as the penalty
increases across the sensitivity range. The five boundaries that remain at
the largest penalty occur around 2026-04-08, 2026-05-07, 2026-05-11,
2026-05-26, and 2026-05-30 UTC.

\begin{figure}[ht]
\centering
\includegraphics[width=1\textwidth]{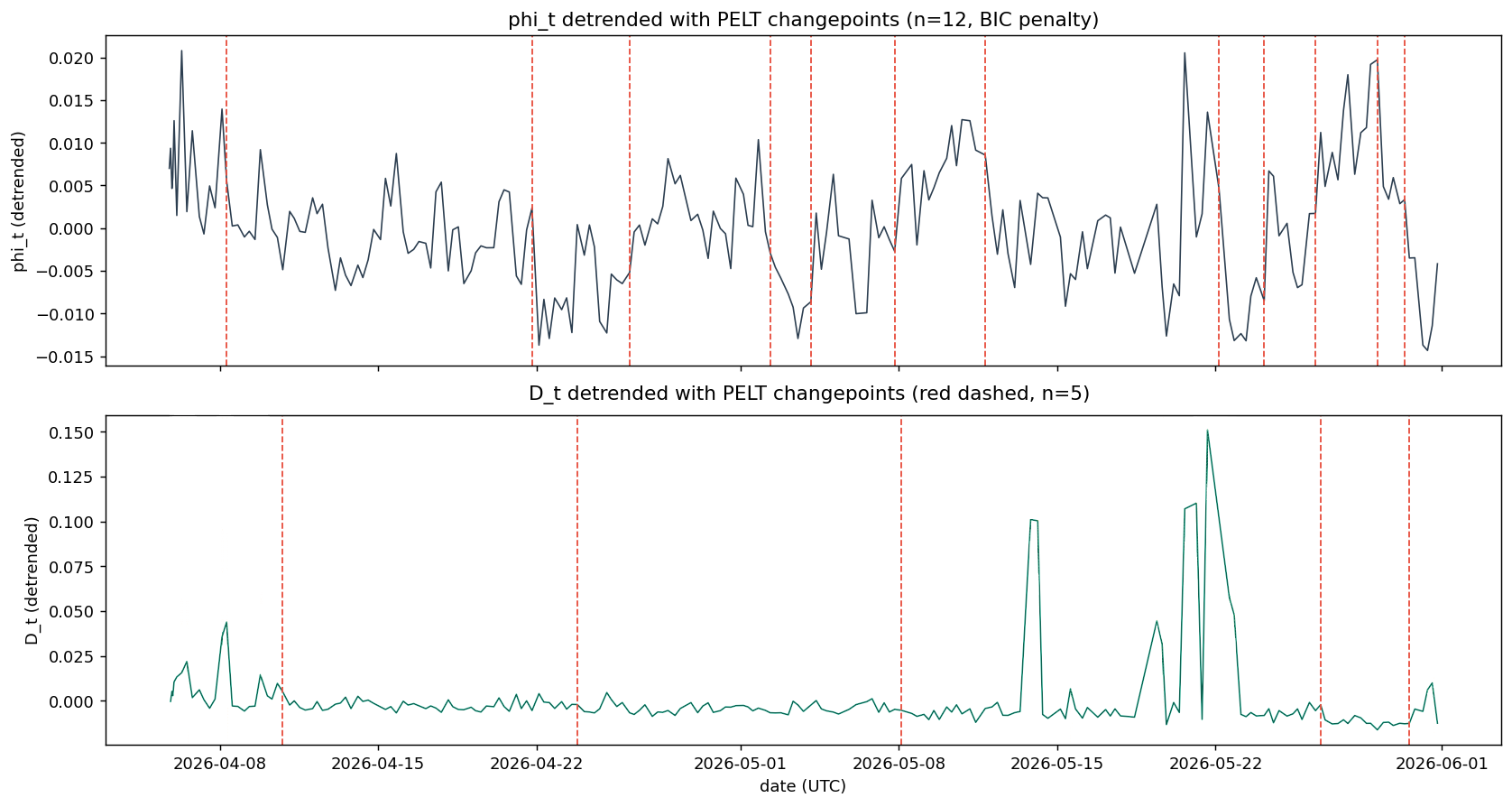}
\caption{Detrended semantic homogeneity $\phit$ and run-to-run drift $\Dt$ with PELT changepoints. Dashed vertical lines mark the boundaries selected at the chosen penalty. The resulting intervals show where the local statistical behavior of each measured series changes.}
\label{fig:phase3-changepoints}
\end{figure}

The detrended $\Dt$ series contains $5$ changepoints at the selected penalty,
with counts of $17$, $5$, and $3$ across the same sensitivity range. The three
boundaries that remain at the largest penalty occur around 2026-04-10,
2026-05-08, and 2026-05-26 UTC. Table~\ref{tab:phase3-changepoints} reports
these boundaries and their penalty sensitivity. Figure~\ref{fig:phase3-changepoints}
shows the full segmentation selected for both series.

\begin{table}[h]
\centering
\small
\setlength{\tabcolsep}{6pt}
\caption{PELT changepoints in the detrended semantic homogeneity $\phit$ and run-to-run drift $\Dt$ series. The sensitivity counts give the number of changepoints at the selected penalty divided by $\sqrt{3}$, at the selected penalty, and at the selected penalty multiplied by $\sqrt{3}$. The listed dates are the boundaries that remain at the largest penalty.}
\label{tab:phase3-changepoints}
\begin{adjustbox}{max width=\textwidth}
\begin{tabular}{lll}
\toprule
Series &
Changepoint count across penalties &
Boundaries remaining at the largest penalty (UTC) \\
\midrule
$\phit$ detrended
&
$22 / 12 / 5$
&
2026-04-08, 2026-05-07, 2026-05-11, 2026-05-26, 2026-05-30
\\
$\Dt$ detrended
&
$17 / 5 / 3$
&
2026-04-10, 2026-05-08, 2026-05-26
\\
\bottomrule
\end{tabular}
\end{adjustbox}
\end{table}

Within this segmentation procedure, the detrended $\phit$ series supports more candidate boundaries than the detrended $\Dt$ series. This indicates that the local statistical behavior of within-run semantic homogeneity changes more often than that of run-to-run drift over the observation window. The boundaries that remain at the largest penalty are the strongest candidates because they persist when additional segmentation carries a greater cost.

These changepoints locate intervals across which the local behavior of the measured series changes. Interpreting a candidate boundary as a transition in collective semantic state would require an accompanying change in another semantic measurement, such as lexical state or cross-layer geometry. Several observations on each side would be needed to distinguish the local state before the boundary from the local state after it. The changepoint analysis identifies where that comparison should be made.

%%%%%%%%%%%%%%%%%%%%%%%%%%%%%%%%%%%%%%%
%%%%%%%%%%%%%%%%%%%%%%%%%%%%%%%%%%%%%%%

\section{Time scales of collective semantic change}\label{sec:temporal-memory}

This section tests whether the sampled feed carries measurable traces of its own recent past. Here, temporal memory means measurable persistence in the sampled feed. Second-order memory refers to persistence through information left in the shared environment, so that earlier feed content may influence later content even when individual agents have no direct memory of earlier posts. The present diagnostics can detect temporal persistence. However, they do not yet identify whether it arises from second-order memory, shared external context, calendar timing, or mechanical effects.

The instrument tests whether the feed behaves as a sequence of mostly independent samples or shows temporal continuity beyond random variation. 
Temporal persistence is read at two levels. Aggregate run-level scalars summarize how quickly measurements of the whole run lose similarity to their own recent past. Layer centroids are read before and after centering controls, which separate the shared static direction of the embedding space from the drifting component. Whether temporal similarity fully decays is the main interpretive question. A remainder at long lag can come from ordinary sources, including drift across months, weekly recurrence, diurnal rhythm, shared static geometry, and overlap between neighboring samples. The section evaluates each of these possibilities before interpreting the remaining temporal structure.

%===================================

\subsection{Aggregate scalar series}
\label{sec:mem-aggregate}

This subsection follows three scalar summaries of the sampled feed across successive runs and examines how quickly each becomes less related to its recent past. The series are semantic homogeneity $\phit$, run-to-run drift $\Dt$, and the consecutive lexical cosine step. Each series is analyzed with two estimators, an autoregressive model of order one, AR(1), and an exponential fit. This gives six decay time estimates $\tau$ from the linearly detrended series. All six point estimates are below $24$ h, the deepest sampled feed age.

The comparison uses linearly detrended versions of all three series. The fitted linear trend is removed before temporal dependence is estimated. The autocorrelation function (ACF) compares a series with itself at lag $k$, where $k$ is the separation in observation steps \citep{box2015time}. The AR(1) model estimates persistence from lag $1$ and converts it to a decay time $\tau$ using the sampling interval \citep{box2015time}. The exponential fit estimates $\tau$ from several lags. The decay time $\tau$ is the time over which the fitted exponential component falls to $e^{-1}$, about $37\%$, of its initial amplitude. The coefficient of determination $R^2$ measures how closely the fitted curve follows the observed ACF \citep{draper1998applied}.

The 95\% confidence intervals use a moving block bootstrap that preserves adjacency and does not wrap the series circularly \citep{kunsch1989jackknife}. Each block contains $24$ grid points, corresponding to six days on the $6$ h grid, and the calculation uses $5000$ resamples. Lagged pairs are formed only within the same block.

\begin{figure}[h]
\centering
\includegraphics[width=1\textwidth]{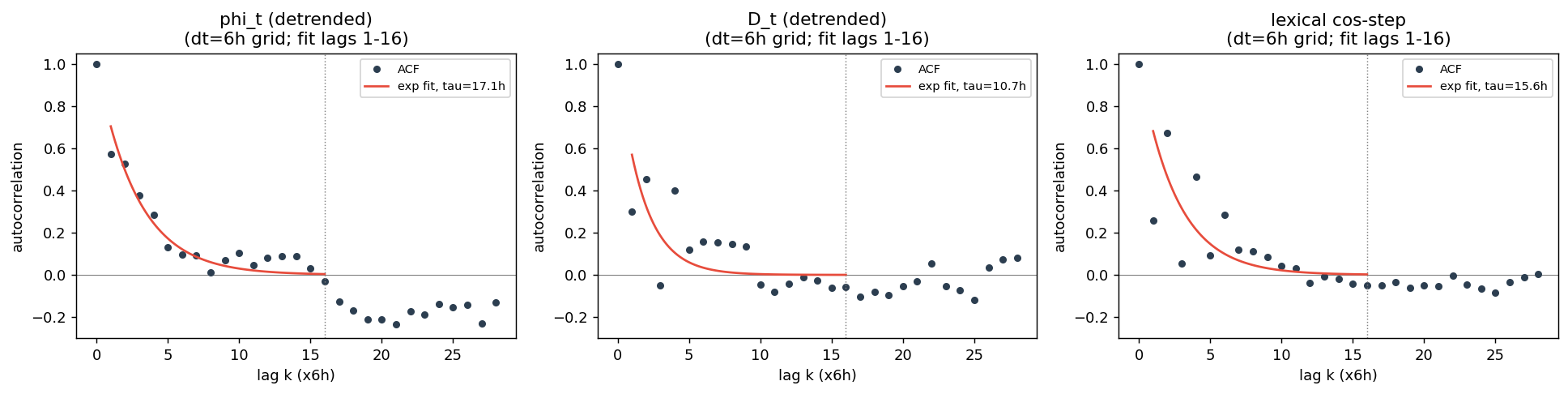}
\caption{Autocorrelation functions and fitted decays (red curves) for the detrended aggregate scalars $\phit$, $\Dt$, and the consecutive lexical cosine step. Exponential decays with zero floor are fitted over lags $1$-$16$ on the $6$ h grid and are shown only over this fit interval. They estimate the decay time of the early positive component.}
\label{fig:phase3-acf-fits}
\end{figure}

For semantic homogeneity $\phit$, the detrended series is used because the raw series has a strong negative trend. The AR(1) estimate is $\tau\approx 10.8$ h, and the exponential fit gives $\tau\approx 17.1$ h with $R^2\approx 0.90$. The corresponding estimates range from $5.0$ to $10.7$ h for run-to-run drift $\Dt$ and from $4.5$ to $15.6$ h for the consecutive lexical cosine step, defined as the cosine distance between the stored top $200$ unigram count vectors in consecutive state rows. All six point estimates indicate a fast component of temporal persistence. The confidence intervals from the exponential fits for $\phit$ and $\Dt$ extend slightly above $24$ h, leaving their exact decay times less precisely resolved. Table~\ref{tab:phase3-memory-timescales} summarizes these decay time estimates. Figure~\ref{fig:phase3-acf-fits} shows the corresponding autocorrelation functions and fitted exponential decays.

\begin{table}[h]
\centering
\small
\setlength{\tabcolsep}{4pt}
\caption{Estimated decay times $\tau$ for the aggregate scalar series. The reported values use linearly detrended series, except for the raw $\phit$ exponential fit, which is included only to show the effect of the strong trend. Its fit has $R^2<0$ and is not interpreted. The $\tau/(24\,\mathrm{h})$ column compares each point estimate with the deepest sampled feed age of $24$ h.}
\label{tab:phase3-memory-timescales}
\begin{adjustbox}{max width=\textwidth}
\begin{tabular}{lllrrr}
\toprule
Series & Detrended & Method & $\tau$ (h) & 95\% CI (h) & $\tau/(24\,\mathrm{h})$ \\
\midrule
$\phit$ & yes & exp fit (lags 1-16) & 17.1 & [7.6, 26.5] & 0.71 \\
$\phit$ & yes & AR(1) & 10.8 & [6.7, 13.5] & 0.45 \\
$\phit$ & raw & exp fit ($R^2<0$, unreliable) & 60.9 & n/a & 2.54 \\
$\Dt$ & yes & exp fit (lags 1-16) & 10.7 & [5.6, 27.3] & 0.44 \\
$\Dt$ & yes & AR(1) & 5.0 & [1.7, 14.8] & 0.21 \\
lexical cosine step & yes & exp fit (lags 1-16) & 15.6 & [5.7, 23.9] & 0.65 \\
lexical cosine step & yes & AR(1) & 4.5 & [1.4, 14.0] & 0.19 \\
\bottomrule
\end{tabular}
\end{adjustbox}
\end{table}

The raw $\phit$ exponential fit gives $\tau\approx 60.9$ h. Its negative $R^2$ means that the exponential curve follows the observed ACF less well than a constant mean. The strong negative trend keeps the raw ACF from decaying toward zero. This value is reported only to show why detrending is required.

In summary, the detrended aggregate scalar series contain autocorrelation with decay time point estimates from about $4.5$ to $17.1$ h. Values of the same scalar become progressively less related as the separation between runs increases. The exponential fits for $\phit$ and $\Dt$ remain imprecise because their confidence intervals extend above $24$ h. Therefore, the scalar analyses resolve a fast temporal component on the scale of hours. A slower component would require persistence in the layer geometry beyond this initial decay, which the next analyses test.

%===================================

\subsection{Raw layer centroid similarity}
\label{sec:mem-raw-layer}

This subsection asks whether the average embedding direction of each post age layer remains similar from one run to the next before the common embedding direction is removed. Removing this common direction is called centering. Raw refers to the original centroid before centering.

For each layer, surface, mid-stream, and residue, the raw layer centroid similarity at lag $1$, $\rho_{1,\mathrm{raw}}$, is the mean cosine similarity between centroids in consecutive usable runs \citep{reimers2019sentencebert}. Cosine similarity measures the alignment of two vectors. A value of $1$ means that they point in the same direction, a value of $0$ means that they have no directional alignment, and a value of $-1$ means that they point in opposite directions.

A high similarity can arise from several sources. It may indicate that related topics, claims, vocabulary, or narrative templates continue to appear in the same layer. It may also reflect repeated mechanical material. In the sampled Moltbook feed, minting posts often repeat nearly identical text across many runs and can help keep the average embedding direction stable. Therefore, the centroid describes the combined composition of the layer, including both discursive posts and repeated mechanical content. It does not identify which component produced the similarity or show that a conversation, argument, or meaning continued from one run to the next.

The raw lag 1 similarities are $0.9565$ for the surface layer, $0.9570$ for the mid-stream layer, and $0.9576$ for the residue layer. Their closeness to $1$ shows that the mean embedding direction of each layer changes very little over the median $5.74$ h separation between consecutive runs.

To describe how this similarity changes across longer separations, the raw cosine similarities over lags $1$-$16$ are fitted with
\begin{equation}\label{eq:layer-centroid-exp}
\rho_k=A e^{-k\,\Delta t/\tau} + C ,
\end{equation}
where $k$ is the lag number, $\Delta t=5.74$ h is the median sampling interval, $A$ is the initial amplitude of the decaying component, $C$ is the fitted offset, and $\tau$ is the decay time. 

The fitted decay time point estimates are $177.1$ h for the surface layer, $165.8$ h for the mid-stream layer, and $193.2$ h for the residue layer. These values fall near one week. Whether that timescale is resolved depends on the uncertainty intervals examined next.

\begin{figure}[H]
\centering
\includegraphics[width=0.5\textwidth]{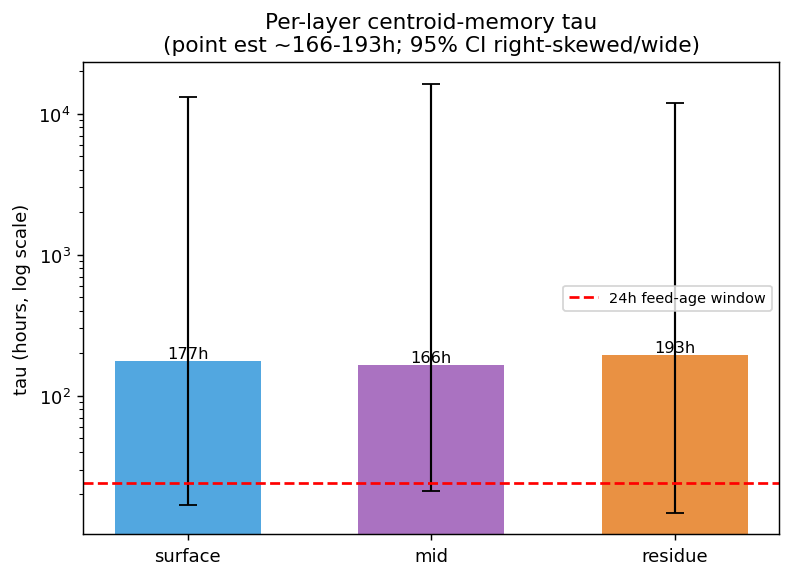}
\caption{Raw centroid decay time estimates for the surface, mid-stream, and residue layers, with moving block bootstrap confidence intervals shown on a logarithmic scale. The intervals use blocks of $24$ adjacent runs and $5000$ resamples. Their lower limits are about $15$ to $21$ h, while their upper limits are roughly $11{,}900$ to $16{,}300$ h, or about $1.4$ to $1.9$ years. Thus, the raw fits do not constrain the decay time to a useful range.}
\label{fig:phase3-perlayer-tau}
\end{figure}

The uncertainty in these fits is extremely large. The lower limits of the $95\%$ confidence intervals range from about $14.8$ to $21.1$ h, while the upper limits range from about $11{,}931$ to $16{,}324$ h, or roughly $1.4$ to $1.9$ years. The bootstrap distributions are strongly skewed toward large values, producing highly asymmetric percentile intervals. These intervals do not constrain the raw layer decay times to a useful range. Figure~\ref{fig:phase3-perlayer-tau} shows the point estimates and confidence intervals on a logarithmic scale, which allows the hour and year scales to be displayed together.

Thus, the raw centroids show very strong similarity between consecutive runs, while the decay times fitted to their longer lag behavior remain unresolved. The observed high raw similarity may reflect continuing themes, repeated vocabulary and templates, persistent minting traffic, and the common direction shared by many embeddings. The next subsection removes that common direction through centering and examines the temporal behavior of the remaining variation.

%===================================

\subsection{Centering and the fast dynamic component}
\label{sec:mem-centering}

The raw layer centroids remain highly similar across runs, although part of this similarity may come from a common direction in the embedding space. This subsection removes that common direction and examines how quickly the remaining variation loses similarity across runs. 

Subtracting a reference vector from every centroid is called centering. The primary analysis uses the grand mean embedding over the two months as the reference vector. Additional checks use separate mean embeddings for April and May or remove a linear trend from each of the $384$ embedding coordinates.

After grand mean centering, lag $1$ cosine similarity $\rho_1$ drops from about $0.96$ to $0.36$-$0.39$. With centering within each month or linear detrending of each coordinate, $\rho_1$ lies between about $0.26$ and $0.32$. Across all centering variants, the AR(1) decay time point estimates range from about $4.2$ to $6.0$ h, comparable to the median sampling interval of $5.74$ h. Much of the fast loss of similarity occurs within the lag $1$ interval. Resolving the shape or exact timescale of this decline would require observations at shorter intervals, which the current record does not provide.

The AR(1) estimate uses only lag $1$ and describes the fast initial loss of similarity. The exponential fit uses lags $1$-$16$ and is influenced by both the positive similarity at lag $1$ and the positive similarity that remains at longer separations. Therefore, the fitted decay time combines the fast initial decline with the slower remaining component and does not resolve either timescale separately.

Table~\ref{tab:phase3-centered-layer-tau} reports layer centroid temporal similarity before and after grand mean centering. The lag $1$ reduction is defined as
\begin{equation*}
\frac{\rho_{1,\mathrm{raw}}-\rho_{1,\mathrm{centered}}}
{\rho_{1,\mathrm{raw}}}.
\end{equation*}
The exponential fit estimates $\tau_{\exp}$ from Eq.~\eqref{eq:layer-centroid-exp} over lags $1$-$16$. Its confidence intervals use the moving block bootstrap described above, with blocks of $24$ positions and $5000$ resamples. The AR(1) estimate uses only lag $1$, with
\begin{equation}
\tau_{\mathrm{AR1}}=-\frac{\Delta t}{\ln\rho_1},
\end{equation}
and its confidence intervals use blocks of $8$ positions and $1000$ resamples.

\begin{table}[H]
\centering
\small
\setlength{\tabcolsep}{4pt}
\caption{Layer centroid temporal similarity before and after grand mean centering. The grand mean embedding over the two months has norm $0.3191$.}
\label{tab:phase3-centered-layer-tau}
\begin{adjustbox}{max width=\textwidth}
\begin{tabular}{lrrrrrr}
\toprule
Layer
& raw $\rho_1$
& centered $\rho_1$
& lag 1 reduction
& raw $\tau_{\exp}$ (h)
& centered $\tau_{\exp}$ (h) [95\% CI]
& centered $\tau_{\mathrm{AR1}}$ (h) [95\% CI] \\
\midrule
surface
& 0.9565
& 0.3664
& 0.617
& 177.1
& 57.4 [22.6, 220.5]
& 5.72 [4.98, 6.36] \\
mid-stream
& 0.9570
& 0.3612
& 0.623
& 165.8
& 100.9 [33.3, 525.9]
& 5.64 [4.93, 6.30] \\
residue
& 0.9576
& 0.3861
& 0.597
& 193.2
& 64.5 [25.5, 158.2]
& 6.03 [5.31, 6.75] \\
\bottomrule
\end{tabular}
\end{adjustbox}
\end{table}

\begin{figure}[h]
\centering
\includegraphics[width=1\textwidth]{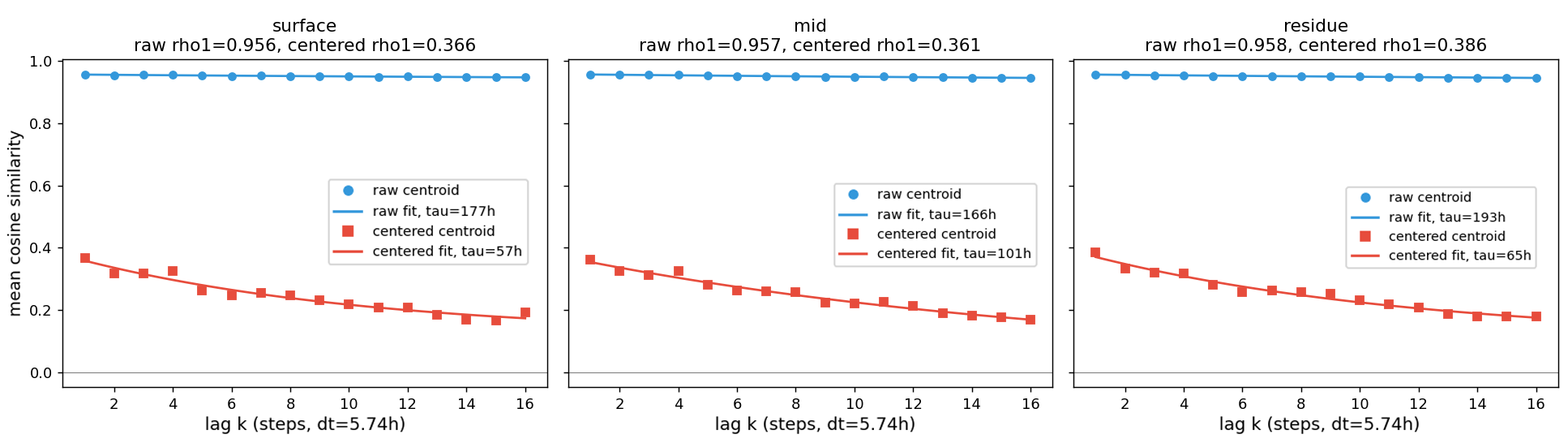}
\caption{Raw and grand mean centered layer centroid cosine similarity across lags, with fitted decays, shown in one panel for each layer. The raw curves (blue) remain near $0.95$-$0.96$. The centered curves (red) begin near $0.36$-$0.39$, decline rapidly, and remain positive over the longest sampled lags.}
\label{fig:phase3-centered-acf}
\end{figure}

For each centering variant, the empirical floor is the mean cosine similarity over lags $12$-$16$. After grand mean centering, the fitted offset $C$ lies below this empirical floor for every layer, with the largest difference in the mid-stream layer. The following analyses consequently use the empirical floor to measure the similarity remaining at longer separations.
Figure~\ref{fig:phase3-centered-acf} shows the corresponding layer centroid cosine similarity across lags.

In summary, centering shows that much of the strong raw similarity comes from the common direction in the embedding space. After this direction is removed, the remaining variation contains a fast temporal component, with a decay time comparable to one sampling interval, although its precise timescale is weakly resolved. A smaller positive similarity remains at longer separations. After grand mean centering, April and May can still lie around slightly different residual centers, which can sustain positive similarity at long lags. Linear drift across the observation window can do the same. The next subsection examines how much of this long lag remainder survives when those broad structures are removed.

%===================================

\subsection{Long lag floor decomposition}
\label{sec:mem-floor}

This subsection asks what remains after the rapid loss of similarity between nearby runs. The floor is defined here as the mean centered layer centroid similarity over lags $12$-$16$. At these separations, the fast component of temporal similarity has largely weakened. A positive floor means that centroid directions separated by many runs still point partly in the same direction on average. At this stage, a positive floor shows similarity at longer separations. A distinct slow temporal component becomes supported only if part of this similarity survives controls for broad structure across the observation window. Here, the word floor refers only to the mean over lags $12$-$16$ and does not assume that the curve remains flat indefinitely. The decomposition tests how much of this long lag similarity is associated with the difference between April and May and with gradual drift over the full observation window.

Grand mean centering uses one common reference for both months and gives a floor of about $0.185$ in all three layers. Across the block bootstrap resamples, all three $95\%$ confidence intervals remain above zero, with lower bounds from $0.115$ to $0.119$. Thus, the positive alignment at these lags remains resolved under block resampling of consecutive runs. Two stronger controls then test whether broad changes over the observation window account for it. Centering by month subtracts separate April and May means, and linear detrending subtracts a fitted linear trend from each embedding coordinate.

Both controls reduce the floor to values from $0.067$ to $0.092$. Every $95\%$ confidence interval remains above zero, with lower bounds from $0.030$ to $0.046$. The floor ratios, measured relative to grand mean centering in the same layer, range from $0.36$ to $0.50$, falling in the middle band between the thresholds $0.3$ and $0.7$ specified before the analysis. The stronger controls remove a substantial part of the similarity at lags $12$-$16$ and leave a smaller positive component in every layer.

The controls remove about $50\%$-$64\%$ of the grand mean centered floor, showing that broad structure across the observation window contributes substantially to the similarity at these lags. A smaller positive remainder survives each control. After centering separately by month, omitting pairs that span April and May raises the floor by $0.005$ to $0.010$, equal to about $3\%$ to $5\%$ of the grand mean floor. If pairs spanning the month boundary produced the floor, removing them would lower it. The observed increase argues against cross-month pairs as the source of the remaining positive similarity. The controls also reduce the lag $1$ cosine similarity $\rho_1$ and the AR(1) decay time $\tau_{\mathrm{AR1}}$. The estimated decay time nevertheless remains on a scale of several hours in every variant. 

\begin{figure}[H]
\centering
\includegraphics[width=1\textwidth]{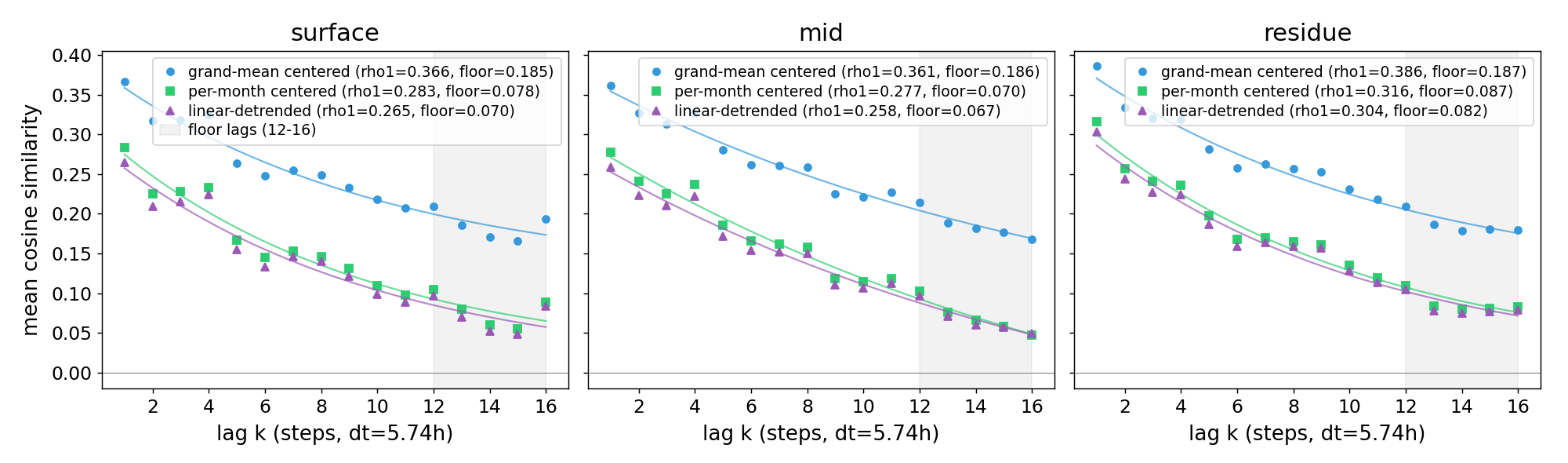}
\caption{Centered cosine ACFs for the three layers under grand mean centering, centering by month, and linear detrending. The shaded region marks the floor interval over lags $12$ through $16$. The stronger controls lower the floor in every layer, but all resulting floor estimates remain positive.}
\label{fig:phase3-floor-acf}
\end{figure}

\begin{table}[H]
\centering
\small
\setlength{\tabcolsep}{4pt}
\caption{Decomposition of the centered long lag floor. The floor is the mean centered cosine ACF over lags $12$-$16$. Grand mean centering is compared with centering by month using all pairs, centering by month using only within-month pairs, and linear detrending of each embedding coordinate. The floor ratio is relative to grand mean centering in the same layer. The thresholds $0.3$ and $0.7$ were specified before the analysis.}
\label{tab:phase3-floor-origin}
\begin{adjustbox}{max width=\textwidth}
\begin{tabular}{llrrrr}
\toprule
Variant & Layer & $\rho_1$ & $\tau_{\mathrm{AR1}}$ (h) & Floor (lags 12-16) [95\% CI] & Floor ratio \\
\midrule
grand mean & surface & 0.366 & 5.72 & 0.1848 [0.115, 0.244] & 1.000 \\
grand mean & mid & 0.361 & 5.64 & 0.1859 [0.119, 0.241] & 1.000 \\
grand mean & residue & 0.386 & 6.03 & 0.1868 [0.116, 0.239] & 1.000 \\
month means, all pairs & surface & 0.283 & 4.55 & 0.0775 [0.044, 0.104] & 0.420 \\
month means, all pairs & mid & 0.277 & 4.47 & 0.0698 [0.032, 0.096] & 0.376 \\
month means, all pairs & residue & 0.316 & 4.98 & 0.0873 [0.044, 0.117] & 0.468 \\
month means, within-month pairs & surface & 0.283 & 4.55 & 0.0837 [0.046, 0.116] & 0.453 \\
month means, within-month pairs & mid & 0.278 & 4.49 & 0.0796 [0.040, 0.111] & 0.428 \\
month means, within-month pairs & residue & 0.316 & 4.98 & 0.0924 [0.046, 0.129] & 0.495 \\
linear detrending & surface & 0.265 & 4.32 & 0.0701 [0.035, 0.104] & 0.379 \\
linear detrending & mid & 0.258 & 4.24 & 0.0668 [0.030, 0.099] & 0.359 \\
linear detrending & residue & 0.304 & 4.82 & 0.0824 [0.039, 0.116] & 0.441 \\
\bottomrule
\end{tabular}
\end{adjustbox}
\par\vspace{0.35em}
\parbox{\textwidth}{\centering\footnotesize\emph{Note.} Confidence intervals come from the block bootstrap.}
\end{table}

Table~\ref{tab:phase3-floor-origin} reports the floor estimates, confidence intervals, floor ratios, lag $1$ similarities, and AR(1) decay times. Figure~\ref{fig:phase3-floor-acf} shows how the ACF changes under each control and marks the interval over which the floor is measured.

In summary, similarity between nearby runs weakens over several hours, while a smaller positive component remains at lags $12$-$16$. Month centering and linear detrending reduce this long lag similarity by about $50\%$-$64\%$, showing that broad structure across the observation window contributes substantially to the grand mean centered floor. A smaller positive remainder survives both controls, supporting a slower temporal component beyond the initial several hour decay. Its source remains unresolved. Excluding cross-month pairs raises the floor slightly, which argues against those pairings as the source of the residual similarity. The next subsection tests whether this surviving component rises again at calendar separations or continues to decline.

%===================================

\subsection{Periodicity check}
\label{sec:mem-periodicity}

This subsection tests whether the slower similarity identified in the layer centroid geometry rises again at daily or weekly separations or continues to decline. Periodic structure appears in an ACF as renewed similarity near the corresponding lag \citep{box2015time}. A weekly recurrence strong enough to account for the long lag layer centroid similarity would produce a rise near one week.

\begin{figure}[H]
\centering
\includegraphics[width=1\textwidth]{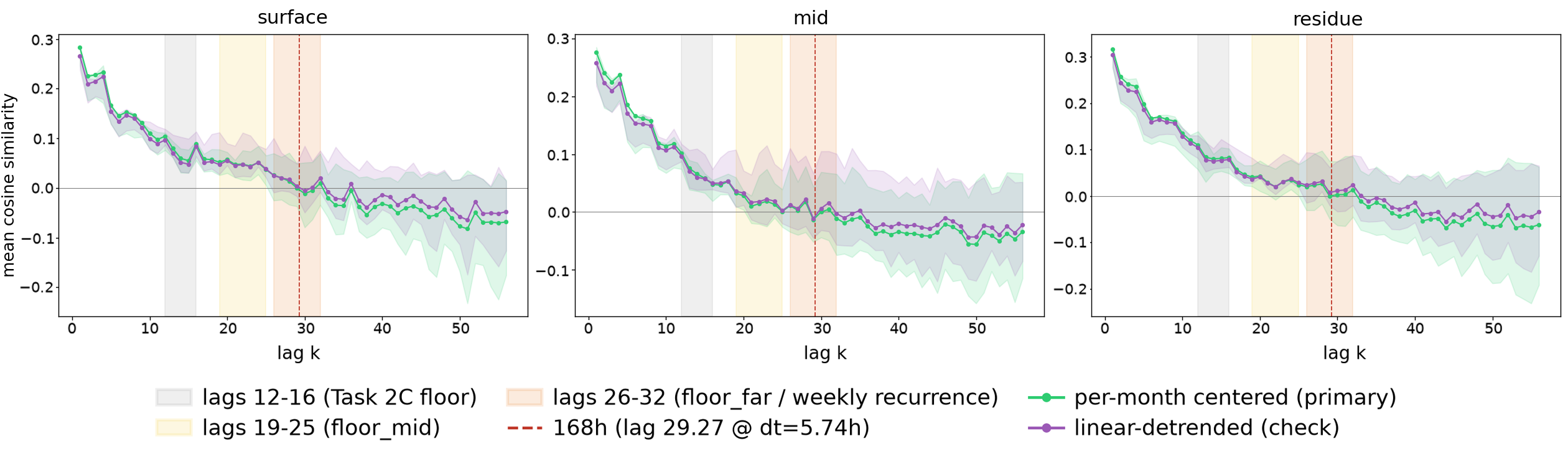}
\caption{Centered layer centroid ACFs through lag $56$, with $95\%$ confidence bands. Shading marks lags $12$-$16$, $19$-$25$, and $26$-$32$. The vertical marker gives the expected weekly lag. Under both stronger centering controls, similarity continues to decrease through the weekly band.}
\label{fig:phase3-longlag-acf}
\end{figure}

The layer centroid ACF is extended from lag $16$ to lag $56$, corresponding to approximately $13.4$ days. Three lag bands are compared. The first is the existing floor over lags $12$-$16$. The second covers lags $19$-$25$. The third covers lags $26$-$32$ and contains the expected weekly lag, $168\,\mathrm{h}/\Delta t \approx 29.27$. Each band value is the mean layer centroid similarity $\rho_k$ across the indicated lags. Figure~\ref{fig:phase3-longlag-acf} shows the full ACF through lag $56$, and Table~\ref{tab:phase3-floor-periodicity} reports the mean similarities and confidence intervals for the three bands.

\begin{table}[H]
\centering
\small
\setlength{\tabcolsep}{4pt}
\caption{Periodicity check for the similarity that remains after centering. The layer centroid ACF is extended to lag $56$, approximately $13.4$ days, and summarized over lags $12$-$16$, $19$-$25$, and $26$-$32$. The last band contains the expected weekly lag, $168\,\mathrm{h}/\Delta t \approx 29.27$. Each entry is the mean $\rho_k$ over the indicated band. A weekly recurrence strong enough to explain the lags $12$-$16$ floor would raise the mean similarity near lag $29$.}
\label{tab:phase3-floor-periodicity}
\begin{adjustbox}{max width=\textwidth}
\begin{tabular}{llrrr}
\toprule
Variant & Layer & $\langle\rho_k\rangle_{12\text{-}16}$ [95\% CI] & $\langle\rho_k\rangle_{19\text{-}25}$ [95\% CI] & $\langle\rho_k\rangle_{26\text{-}32}$ [95\% CI] \\
\midrule
centered by month, all pairs & surface & 0.0775 $[0.055, 0.089]$ & 0.0480 $[0.026, 0.063]$ & 0.0074 $[-0.041, 0.055]$ \\
centered by month, all pairs & mid & 0.0698 $[0.047, 0.076]$ & 0.0164 $[-0.010, 0.044]$ & 0.0013 $[-0.045, 0.058]$ \\
centered by month, all pairs & residue & 0.0873 $[0.050, 0.109]$ & 0.0316 $[0.011, 0.068]$ & 0.0125 $[-0.043, 0.079]$ \\
linear detrending & surface & 0.0701 $[0.057, 0.106]$ & 0.0468 $[0.033, 0.094]$ & 0.0116 $[-0.014, 0.065]$ \\
linear detrending & mid & 0.0668 $[0.052, 0.100]$ & 0.0206 $[-0.006, 0.085]$ & 0.0065 $[-0.035, 0.092]$ \\
linear detrending & residue & 0.0824 $[0.066, 0.113]$ & 0.0315 $[0.022, 0.074]$ & 0.0196 $[-0.016, 0.073]$ \\
\bottomrule
\end{tabular}
\end{adjustbox}
\par\vspace{0.35em}
\parbox{\textwidth}{\footnotesize\emph{Note.} Confidence intervals use a moving block bootstrap, which resamples consecutive groups of runs to preserve local temporal dependence. This lag $56$ analysis uses block length $64$. Therefore, its lags $12$-$16$ intervals differ slightly from those in Table~\ref{tab:phase3-floor-origin}.}
\end{table}

Figure~\ref{fig:phase3-longlag-acf} and Table~\ref{tab:phase3-floor-periodicity} show the same pattern across both centering controls and all three layers. Mean similarity decreases across the three bands,
\begin{equation*}
\left\langle \rho_k \right\rangle_{26\text{-}32}
<
\left\langle \rho_k \right\rangle_{19\text{-}25}
<
\left\langle \rho_k \right\rangle_{12\text{-}16}\,.
\end{equation*}
The mean similarities over lags $12$-$16$ range from $0.0668$ to $0.0873$. They decrease to values from $0.0164$ to $0.0480$ over lags $19$-$25$ and to values from $0.0013$ to $0.0196$ over lags $26$-$32$. Every confidence interval for the lags $12$-$16$ band remains above zero, while every confidence interval for the weekly band includes zero. Thus, the remaining similarity continues to decline through the expected weekly lag, with no rise consistent with weekly recurrence.

A separate diurnal comparison groups run pairs by the difference between their times of day. Pairs whose times of day differ by $0$-$6$ hours have mildly higher layer centroid similarity than pairs whose times differ by $12$-$18$ hours. The size of this asymmetry is too small to account for the positive mean similarity over lags $12$-$16$. The observation window contains about eight weekly cycles, which limits sensitivity to a weak weekly pattern. Together with the continued decline in Figure~\ref{fig:phase3-longlag-acf}, this comparison gives no evidence that daily or weekly recurrence accounts for the remaining long lag similarity.

%===================================

\subsection{Extended temporal structure of semantic homogeneity}
\label{sec:phi-extended-temporal}

In the April-May analysis, the $\phit$ ACF was calculated after a single linear trend was removed across the full two-month interval. It decayed from positive values at short lags and formed a broad negative lobe over lags $19$-$25$. The mean ACF in this band was $-0.1604$, with a $95\%$ CI of $[-0.2972,-0.0282]$. At larger lags the ACF rose again toward zero, giving the two-month curve an apparent oscillatory form. The short record did not resolve whether this structure persisted or whether its apparent timescale was stable.

The negative lobe may depend on how the broad variation in $\phit$ is treated. We repeat the calculation on the same April-May observations with within-month centering. The mean value of $\phit$ is calculated separately for April and for May, the April mean is subtracted from every April observation, and the May mean is subtracted from every May observation. The ACF recomputed from this centered series describes fluctuations of $\phit$ around the level measured within each month.

After centering by month, the mean over the same lags is $-0.0033$, with a $95\%$ CI of $[-0.1470,$ $0.0417]$. The grid, ACF estimator, bootstrap procedure, and random seed are unchanged. The disappearance of the negative lobe under monthly centering shows that the apparent oscillatory form of the April-May ACF depends on how the broad variation in $\phit$ is removed.

Observations continued after the April-May analysis, providing approximately two additional months of data through early August. In this extended series, the monthly mean of $\phit$ decreases from $0.1106$ in April to $0.1011$ in May and $0.0946$ in June, then rises to $0.1165$ in July and $0.1228$ in the partial August record. These changes remain part of the measured temporal structure. With only five monthly estimates, and with August incomplete, the broad component does not support a resolved temporal timescale.
To examine shorter temporal dependence in the extended data, we center $\phit$ within each calendar month and compute the ACF of the remaining fluctuations. This operation defines a local monthly level without assigning a functional form to the broader variation.

After centering by month, the extended $\phit$ ACF decreases from $0.7559$ at lag $1$ to $0.1684$ at lag $15$. The $95\%$ confidence intervals exclude zero through lag $15$, corresponding to approximately $3.75$ days on the $6$ h grid. At lag $16$, or four days, the ACF is $0.1083$ with a confidence interval that includes zero, and every confidence interval from lag $16$ through lag $56$ also includes zero. The loss of resolved dependence occurs well within the monthly interval over which the reference mean is defined. Thus, within-month fluctuations in $\phit$ show resolved temporal dependence over several days, with no resolved dependence beyond approximately four days in the centered series.

\begin{figure}[h]
\centering
\includegraphics[width=0.68\textwidth]{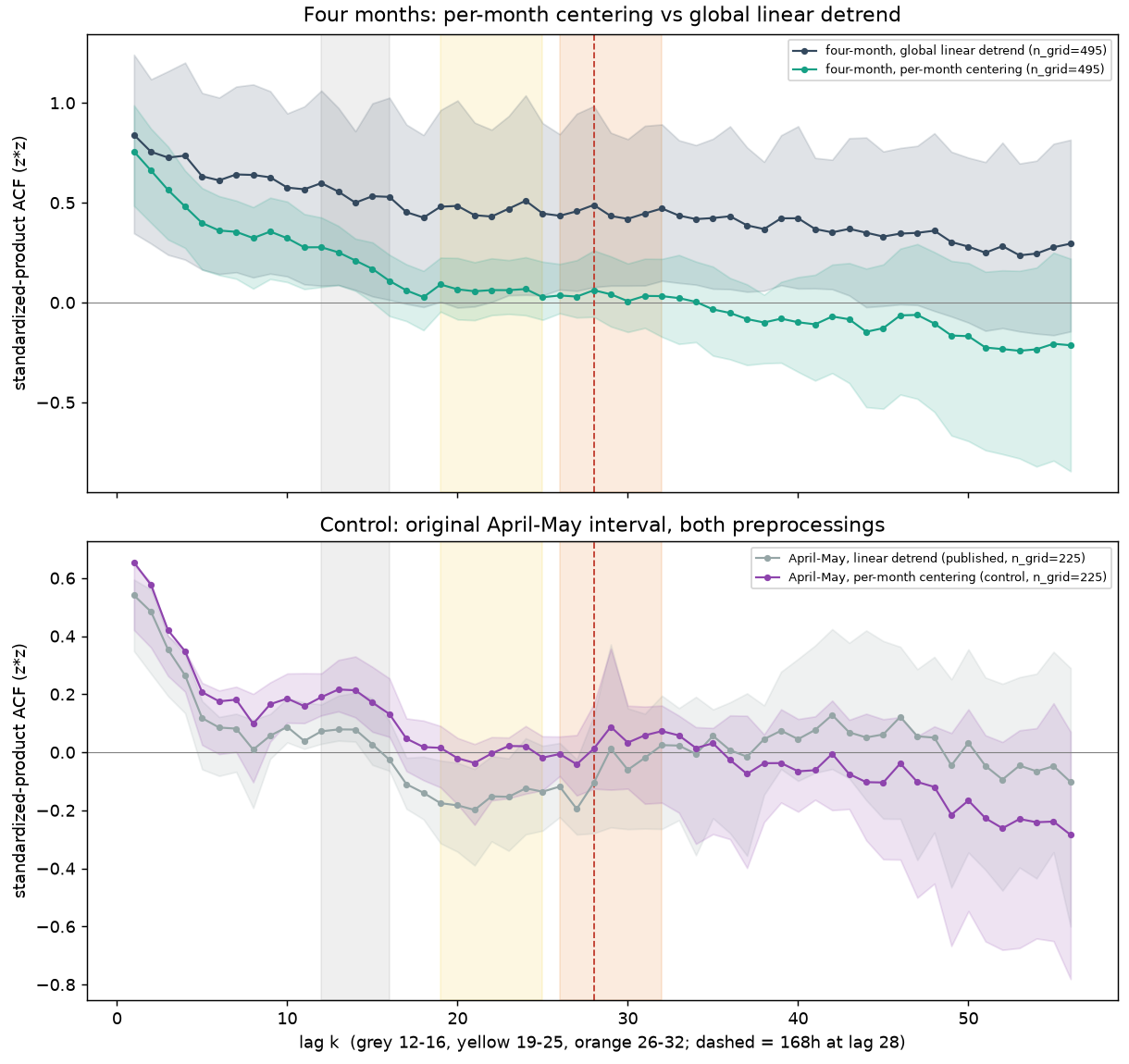}
\caption{Sensitivity of the $\phit$ long lag ACF to preprocessing and observation duration. The lower panel compares linear detrending with centering by month on the same April-May series. The negative lobe over lags $19$-$25$ disappears under centering by month. The upper panel extends the comparison through early August. Centering within each calendar month removes changes in monthly mean from the ACF calculation, leaving fluctuations around the mean within each month. Their temporal dependence remains resolved through lag $15$, corresponding to approximately $3.75$ days on the $6$ h grid, with no distinct daily or weekly recurrence peak.}
\label{fig:phase3-phi-periodicity-control}
\end{figure}

The expected weekly separation occurs at lag $28$ on the $6$ h grid. At this lag the ACF is $0.0614$, with a $95\%$ CI of $[-0.0721,0.2608]$. The mean ACF over lags $26$-$32$ is $0.0341$, with a $95\%$ CI of $[-0.0992,0.1931]$. No renewed rise is resolved near one week. The daily separation occurs at lag $4$ within the initial decay and does not form a distinct recurrence peak. Figure~\ref{fig:phase3-phi-periodicity-control} summarizes both controls, comparing linear detrending with centering by month on the April-May record and extending the same comparison through early August.

In summary, the extended $\phit$ record resolves temporal structure on two scales. The monthly mean changes across the observation period, though the available five monthly estimates do not resolve a timescale for this broader variation. After centering within each month, fluctuations in $\phit$ remain temporally related for approximately four days and show no distinct daily or weekly recurrence. The negative lobe seen in the original April-May ACF does not survive the preprocessing control and is not assigned an oscillatory interpretation. Resolving the timescale of the broader component would require a longer record containing enough monthly variation to distinguish persistent change from recurrent structure.

%%%%%%%%%%%%%%%%%%%%%%%%%%%%%%%%%%%%%%%
%%%%%%%%%%%%%%%%%%%%%%%%%%%%%%%%%%%%%%%

\section{Discussion}
\label{sec:discussion}

The object measured by our instrument is the semantic state of an agent population. At each observation, the sampled texts occupy a distribution of positions in embedding space and a distribution of lexical content. The run mean embedding gives the central semantic position of that population sample, semantic homogeneity measures average alignment among its embedded texts, and the post age layers resolve how different temporal parts of the sampled population are positioned relative to the same run state. Repeated observations turn these population quantities into trajectories.

A population observable is not a scaled version of an individual one. Semantic homogeneity is defined on a sample of many outputs and has no value for a single output, in the same way that temperature is defined on an ensemble and carries no information about the path of one molecule. The collective semantic center can move while individual outputs stay widely dispersed. Collective alignment can strengthen or weaken with the center held nearly fixed. Different parts of the population can move differently relative to the shared state, and the collective position can follow a trajectory that no individual agent follows. These degrees of freedom belong to the ensemble of observed outputs.

The measurement design follows from this. A run of the instrument is not interpreted as one conversation, and the layer centroids do not reconstruct a path of communication among the sampled agents. Threads may be disconnected, agents may respond to different local histories, and the sampled feed may contain several simultaneous forms of activity. These differences contribute to the population state that the instrument observes. Interaction structure becomes necessary when the question changes from whether the population state moved to how that motion was produced.

The present study shows that these population level degrees of freedom are measurable. The collective semantic center moves between observations, alignment among sampled outputs changes independently of that motion, the three post age layers move substantially relative to the shared run state, and several population quantities remain related to their own recent values across successive runs. Our instrument resolves collective semantic motion without requiring that the sampled population form a single conversation or that the mechanism producing that motion already be known.

The mixed feed defines the population measured in the present study. Discursive posts, repeated minting traffic, promotional material, and other content returned under the observation protocol all contribute to the measured semantic state. The scalar, lexical, and embedding quantities describe changes in this mixture. All measurements derived from post content use only the stored text, consisting of the title and the first $300$ characters of each fetched post, and their interpretation applies to that sampled text.

Temporal similarity alone cannot identify why nearby observations resemble one another. Second-order memory is one possible source. Under this mechanism, information left in the shared environment allows earlier feed content to influence later content. Shared exogenous context is the principal rival explanation. Several observations may resemble one another because agents respond to the same news cycle, platform event, task, or other condition originating outside the sampled feed. This mechanism can produce temporal dependence without influence from earlier sampled content. The shared temporal context identified by \citet{li2026socialization} provides a population scale example of this possibility on Moltbook.

The present measurements constrain the timescales on which such explanations must operate. Consecutive lexical states are closer than states paired after timestamp shuffling. Detrended scalar quantities and centered layer centroids lose much of their similarity over several hours. After the stronger controls, a smaller mean layer centroid similarity of approximately $0.067$-$0.092$ remains over lags $12$-$16$ and continues to decline at larger separations. No strong rise occurs near the expected weekly lag. A common external driver that changes only over many days cannot by itself account for the rapid loss of similarity over several hours. External context that changes on a comparable timescale remains compatible with the rapid component.

Several sources remain compatible with the slower structure. Repeated returns to similar topics can increase similarity between observations separated in time. Changes in the prevalence of minting or promotional traffic can move the measured state of the mixed feed. Platform drift can change the population from which the feed is sampled. Residual directional structure in the embedding model can preserve positive centroid similarity even after centering. Sampling overlap is a more specific contribution, because consecutive residue windows overlap at the typical sampling interval and can increase residue similarity at lag $1$. Surface and mid-stream windows do not overlap at that interval, overlap between neighboring residue windows cannot account for the positive similarity measured over lags $12$-$16$, and the agreement of the fast component and the longer lag behavior across the three layers further limits an explanation based only on residue overlap. These sources need not act separately and can contribute at different timescales.

The treatment of broad variation before a temporal statistic is computed can also create apparent temporal structure. Section~\ref{sec:phi-extended-temporal} gives a worked case within the present measurements. A single linear trend removed across the two-month interval left a negative lobe in the $\phit$ autocorrelation over lags $19$-$25$, with a mean value of $-0.1604$. Centering within each calendar month, applied to the same observations on the same grid with the same estimator, reduced that mean to $-0.0033$. The apparent oscillatory form depends on the detrending choice. The short range persistence survives both treatments. Any reading of periodicity, recurrence, or oscillation in these series carries its preprocessing specification with it.

The candidate changepoints in Section~\ref{sec:changepoints} create a measurable question that the present quantities can already address. A boundary in a scalar series becomes evidence for a change between dynamical regimes only when the observations on its two sides show persistently different behavior. Local means, variances, temporal decay, lexical state, and cross-layer geometry can be estimated on the observations preceding and following a candidate boundary. A regime interpretation would require the difference to persist across several observations on each side and to appear in an independent measure of the sampled content. A brief excursion followed by a return to the previous behavior would instead remain a transient perturbation. Repeated observation provides the stronger test. If comparable boundaries occur again, the same dynamical quantities can be evaluated around each event, and recurrence of the same transition pattern would support a reproducible regime description. Failure to reproduce it would keep the original boundary at the level of candidate segmentation.

One internal control can be strengthened without any new stored field. The deterministic shuffle comparison currently supplies $13$ usable entries. A shuffle measurement at every observation would produce a time series of the null cross-layer geometry. The observed minus shuffled distances could then be tested for change through time, including whether the null geometry itself follows changes in feed composition.

Four additional records would move the analysis from collective motion itself to the mechanisms that generate it. The present configuration does not store or independently measure these observables.

The first is the individual embeddings of the stored texts from which the layer centroids were formed. A centroid records the mean semantic direction of a sampled layer and discards much of its internal structure. Two layers can have similar centroids while containing different clusters, broad distributions, or multiple semantic groups. Individual embeddings, stored together with content labels for minting, promotional material, and recurring topics, would allow the distribution within each layer to be reconstructed, the embedding quantities to be recomputed on defined content subsets, and persistent multimodal structure to be tested across observations. The difference between the resulting series would measure directly how much a content class contributes to the population level signal.

The second is a set of stable post and agent identifiers. With identifiers present, individual and population trajectories could be compared, and candidate semantic groups could be followed through successive observations. Whether a semantic partition persists across observations, or reforms from different members at each run, becomes a measurable property of the population.

The third is direct links between earlier and later posts. Such links would identify candidate transmission paths and allow distinctive claims, phrases, or semantic structures to be followed through later content. Combined with individual semantic trajectories, they would answer whether a distinctive semantic change appears first in one part of the interaction structure and subsequently along connected paths. Compared with the semantic partition above, they would also answer whether a candidate subgroup coincides with the topology of actual exchanges. Persistence of the same partition across semantic and interaction structure would provide evidence for organized subgroup behavior.

The fourth is an independent record of external events and topics. Such a record would allow the component associated with shared exogenous context to be estimated and the residual series tested again for temporal dependence \citep{draper1998applied,box2015time}. Persistence after that adjustment would strengthen would strengthen the case for second-order memory, while a reduction toward the corresponding shuffle reference would quantify the contribution of the measured common context. Independent records of post volume and feed composition would allow the contribution of platform drift to be tested in the same way.

Moltbook serves as a validation setting for a more general measurement design. The object measured by our instrument is the semantic state of a shared language environment observed repeatedly in time and divided according to the age of the content present at each observation. That construction does not depend on Moltbook specific semantics. It requires timestamped language activity, a defined observation schedule, and a rule for assigning observed content to age layers. In an operational agent team, the corresponding language environment could include shared messages, memory entries, task records, tool mediated text, and operator instructions. The same design would produce a sequence of population states while preserving the distinction between newly introduced content and content that has remained available to the system. Semantic displacement, homogeneity, cross-layer geometry, persistence, recurrence, and candidate regime changes would have the same measurement meaning under the defined operational record.

Operational systems can expose many of these observables as part of their ordinary operation. Task assignments and operator inputs provide independent records of exogenous context. Individual messages and memory entries carry their own embeddings and identifiers. Communication and memory access links provide candidate relay paths. The inference conditions stated above become direct tests within the system being monitored. Controlled experiments can move one step further. A defined semantic perturbation can be introduced into a shared memory or communication channel while task conditions are recorded, and Kopterix can then measure its onset, displacement across content age layers, persistence, recurrence, and recovery relative to matched runs without the perturbation. This turns the observational distinction between external context and second-order memory into an experimental comparison with a known source and a known introduction time.

The central result is that collective semantic change can be treated as a dynamical property of an agent population and measured while that population is operating. The present study resolves motion, alignment, relative layer structure, and temporal persistence at the population level while leaving their mechanisms open where the required observables are absent. Kopterix provides a controlled design for carrying those measurements forward as the available records move from population state toward causal structure.

%%%%%%%%%%%%%%%%%%%%%%%%%%%%%%%%%%%%%%%
%%%%%%%%%%%%%%%%%%%%%%%%%%%%%%%%%%%%%%%

\section{Conclusion}

This study shows that collective semantic change in a population of language model agents can be measured as a dynamical phenomenon. Our instrument, Kopterix, observes that change as a sequence of bounded observations of population state under a protocol defined before the observations begin. Each observation divides the sampled feed by post age into the \surface{}, \midstream{}, and \residue{} layers. This age stratification places recent and older content inside the same run, which makes semantic differences across content age measurable alongside run-to-run change.

At the lexical level, rarefaction reduced the dependence of entropy on sample support and left a clear April-May difference at a fixed token budget. The distribution formed from the stored top $200$ unigram counts was less even in May than in April. Adjacent state rows were also lexically closer than rows paired after timestamp shuffling. These results resolve change in lexical concentration and temporal organization of the sampled token distributions within the present observation window.

At the geometric level, raw layer centroids remained highly aligned across neighboring runs. Grand mean centering removed most of this similarity and exposed the scale of the common embedding direction. In the $13$ usable shuffle checks, the mid-stream and residue layers were farther apart than after shuffling in $10$ cases, with an average excess of $0.0166$. The separations between surface and mid-stream, and between surface and residue, did not show the same consistency. The available control supports a recurring excess in separation between the mid-stream and residue layers within those checks. A claim of stable geometric organization across all three post age layers remains unsupported by the present observation record.

At the temporal level, detrended aggregate quantities and centered layer centroids lose much of their similarity over several hours, near the observation cadence. Stronger centering and detrending leave a weaker positive component at longer separations. That component continues to decline across the extended lag range, and the periodicity checks show no strong weekly recurrence. The raw layer centroid exponential fits produced week scale point estimates. Their confidence intervals span from hours to years and do not resolve a week scale memory time.

Several attractive apparent structures failed the checks required for a stronger reading. The modal cross-layer ordering seen in April did not recur in May. The global layer label permutation check did not distinguish the observed labeling from the other five assignments. The approximately $120^\circ$ residual angle pattern followed from the residual centering construction and could not serve as independent geometric evidence. These failures demonstrate the role of the inference ceiling in the instrument. Each interpretation is limited to the level supported by its controls and by independent observables. Under that ceiling, the present study can measure lexical change, cross-layer geometry, temporal dependence, persistence, recurrence, and candidate segmentation in the sampled population state. It does not identify discursive interaction, second-order memory, or a causal platform mechanism.

Moltbook provides the validation setting for this measurement design. The design applies wherever a population of language model agents produces a timestamped language environment that can be observed repeatedly and divided by content age. Shared messages, memory entries, task records, tool mediated text, and operator instructions can supply such an environment when their timing and population scope are available. The same observational logic can then compare newly introduced, intermediate, and older semantic material while following the population state across observations. Collective semantic change in populations of language model agents is a measurable dynamical phenomenon. Kopterix observes it as it develops.

%%%%%%%%%%%%%%%%%%%%%%%%%%%%%%%%%%%%%%%
%%%%%%%%%%%%%%%%%%%%%%%%%%%%%%%%%%%%%%%

\section{Reproducibility and provenance}

The archived data products, analysis code, software environment, reconstruction records, corrected bootstrap settings, superseded calculations, and known limits on exact numerical reproduction are documented in the accompanying Zenodo deposit \citep{kopterix2026validation}. The archived code assumes nanosecond datetime resolution when timestamps are cast to integers. Current pandas versions may use microsecond resolution, which changes the scale of time based calculations; explicit conversion to nanosecond precision before the cast restores the intended scale. Subject to the reproduction limits documented in the deposit, the archived materials support regeneration of the numerical analyses based on the April-May study window. 

The extended $\phit$ analysis in Section~\ref{sec:phi-extended-temporal} recomputes the April-May autocorrelation under per-month centering and uses observations continued through early August. Both calculations lie beyond the frozen deposit. Corresponding scripts, reduced input table, output tables, and checksums are published in the folder \texttt{extension\_4months/} in the public GitHub mirror of the deposit.\footnote{\url{https://github.com/elena-kopteva/kopterix-deposit}} The folder was added alongside the frozen deposit v1.0.0 contents, which remain unmodified. Verification of the exact deposit files is against the Zenodo archive; the
mirror serves as a reference copy.

OpenAI ChatGPT and Anthropic Claude, including Cowork, assisted with code development, data processing, statistical checks, project audits, and language editing. The author designed the instrument, defined the scientific questions, selected and reviewed the analyses, verified the source data and computational outputs, interpreted the evidence, and takes responsibility for the results reported here.

%%%%%%%%%%%%%%%%%%%%%%%%%%%%%%%%%%%%%%%
%%%%%%%%%%%%%%%%%%%%%%%%%%%%%%%%%%%%%%%

\section*{Acknowledgments}

The author thanks Vitaliy Hlynianyi-Zhuk for his sustained intellectual engagement with the project, many valuable discussions, and crucial feedback throughout its development.

%%%%%%%%%%%%%%%%%%%%%%%%%%%%%%%%%%%%%%%
%%%%%%%%%%%%%%%%%%%%%%%%%%%%%%%%%%%%%%%

\begin{appendices}

\section{Geometric origin of the residual angle baselines}\label{app:residual-angle-geometry}

The residual angle analysis uses two geometric reference values. Independent random directions in the $384$ dimensional embedding space concentrate near $90^\circ$. Three vectors of similar length that approximately sum to zero instead form pairwise angles near $120^\circ$. This appendix derives both results.

\subsection{Independent random directions in high dimensions}

Let $\mathbf u$ and $\mathbf v$ be independent random unit vectors in $\mathbb R^d$. The angle $\theta$ between them is defined by
\begin{equation}
\cos\theta
=
\frac{\mathbf u\cdot\mathbf v}
{\|\mathbf u\|_2\,\|\mathbf v\|_2}.
\end{equation}
Since both vectors have unit norm,
\begin{equation}
\cos\theta
=
\mathbf u\cdot\mathbf v.
\end{equation}

The distribution of a random direction is rotationally symmetric. Therefore, we may rotate the coordinate system so that the first vector points along the first coordinate axis,
\begin{equation}
\mathbf u
=
(1,0,\ldots,0).
\end{equation}
Writing the second vector as
\begin{equation}
\mathbf v
=
(v_1,v_2,\ldots,v_d),
\end{equation}
the cosine of the angle becomes
\begin{equation}
\cos\theta
=
\mathbf u\cdot\mathbf v
=
v_1.
\end{equation}

Because $\mathbf v$ has unit norm, its squared components satisfy
\begin{equation}
v_1^2+v_2^2+\cdots+v_d^2
=
1.
\end{equation}
Since no coordinate direction is preferred, each squared component contributes the same expected share of the total,
\begin{equation}
\mathbb E[v_i^2]
=
\frac{1}{d}.
\end{equation}
The distribution is symmetric around zero, meaning
\begin{equation}
\mathbb E[v_i]
=
0.
\end{equation}
Hence, the standard deviation of each component is
\begin{equation}
\sqrt{\mathbb E[v_i^2]}
=
\frac{1}{\sqrt d}.
\end{equation}

Since $v_1=\cos\theta$, the cosine between two independent random directions has a characteristic scale
\begin{equation}
\cos\theta
\sim
\frac{1}{\sqrt d}
\end{equation}
around zero. For the embedding dimension used in this study,
\begin{equation}
\frac{1}{\sqrt{384}}
\approx
0.051.
\end{equation}

The factor $1/\sqrt{384}$ describes the typical magnitude of the projection of one random unit vector onto any previously chosen direction. A random unit vector spreads its squared magnitude across $384$ coordinate directions, with each coordinate contributing approximately $1/384$ on average.

Therefore, the cosine between two independent random directions is concentrated near zero, and its expected value is zero,
\begin{equation}
\cos\theta
\approx
0,
\end{equation}
which gives
\begin{equation}
\theta
\approx
\arccos(0)
=
\frac{\pi}{2}
=
90^\circ.
\end{equation}

Near $\pi/2$,
\begin{equation}
\arccos x
=
\frac{\pi}{2}-x+O(x^3).
\end{equation}
A cosine scale of $1/\sqrt{384}\approx 0.051$ corresponds to an angular scale of approximately $0.051$ rad, or about $2.9^\circ$, around $90^\circ$.

A standard Gaussian vector has a rotationally symmetric direction. Normalizing such a vector produces a random direction on the unit sphere. Rescaling it afterward to match an observed residual norm changes its length while leaving its direction and all pairwise angles unchanged. Gaussian vectors rescaled to the observed residual norms therefore retain the approximately $90^\circ$ angle baseline.

\subsection{Three centered vectors of equal length}

Now consider three residual vectors that sum to zero,
\begin{equation}\label{eq:app-zero-sum}
 r_S+ r_M+ r_R
=
0.
\end{equation}
Suppose first that they have the same norm,
\begin{equation}
\| r_S\|_2
=
\| r_M\|_2
=
\| r_R\|_2
=
a.
\end{equation}

The zero sum condition in Eq.~\eqref{eq:app-zero-sum} gives
\begin{equation}
 r_R
=
-\left( r_S+ r_M\right).
\end{equation}
Taking the squared norm of both sides,
\begin{equation}
\| r_R\|_2^2
=
\| r_S+ r_M\|_2^2.
\end{equation}
Expanding the right hand side gives
\begin{equation}
\| r_R\|_2^2
=
\| r_S\|_2^2
+
\| r_M\|_2^2
+
2 r_S\cdot r_M.
\end{equation}
Substituting the common norm $a$,
\begin{equation}
a^2
=
a^2+a^2
+
2 r_S\cdot r_M,
\end{equation}
from which it follows that
\begin{equation}
 r_S\cdot r_M
=
-\frac{a^2}{2}.
\end{equation}

The dot product can also be written in terms of the angle $\theta_{SM}$,
\begin{equation}
 r_S\cdot r_M
=
\| r_S\|_2\,
\| r_M\|_2
\cos\theta_{SM}
=
a^2\cos\theta_{SM}.
\end{equation}
Equating the two expressions gives
\begin{equation}
a^2\cos\theta_{SM}
=
-\frac{a^2}{2},
\end{equation}
and hence
\begin{equation}
\cos\theta_{SM}
=
-\frac{1}{2}.
\end{equation}
The angle is therefore
\begin{equation}
\theta_{SM}
=
\arccos\left(-\frac{1}{2}\right)
=
\frac{2\pi}{3}
=
120^\circ.
\end{equation}

The same derivation applies to the other two pairs. Three equal length vectors that sum to zero form a symmetric configuration in which every pair is separated by $120^\circ$. Although the vectors lie in a $384$ dimensional ambient space, this balanced configuration lies in the two dimensional plane spanned by any two of the vectors.

The argument $-1/2$ inside the inverse cosine is the normalized dot product,
\begin{equation}
\frac{
 r_S\cdot r_M
}{
\| r_S\|_2\,\| r_M\|_2
}
=
-\frac{1}{2}.
\end{equation}
Its negative sign indicates an obtuse angle. Its magnitude fixes that angle at $120^\circ$.

\subsection{Approximate centering in the measured residuals}

The measured residuals satisfy
\begin{equation}
 r_S(t)+ r_M(t)+ r_R(t)
=
3\left(\bar c(t)-\bar e(t)\right).
\end{equation}
They form an exact zero sum configuration when the run mean embedding $\bar e(t)$ equals the equal weight centroid mean $\bar c(t)$. In most observations, the measured difference between these two means is very small, and the zero sum geometry is a close approximation.

The three residual norms are also similar in scale across the record. Together, approximate centering and comparable residual norms make $120^\circ$ the natural reference value for the observed angle distribution. Differences among the residual norms and the measured gap between $\bar e(t)$ and $\bar c(t)$ both produce departures from this symmetric value.

The two reference angles, $90^\circ$ and $120^\circ$, describe distinct geometries. Independent random directions in a high dimensional space concentrate near $90^\circ$ because their normalized dot products concentrate near zero. For three residual vectors of similar length, centering constrains the vectors to approximately balance one another, so their pairwise angles approach $120^\circ$. The second pattern is a geometric consequence of the residual construction and provides the baseline against which departures in the measured residual geometry can be evaluated.

\end{appendices}

%%%%%%%%%%%%%%%%%%%%%%%%%%%%%%%%%%%%%%%
%%%%%%%%%%%%%%%%%%%%%%%%%%%%%%%%%%%%%%%

\bibliographystyle{plainnat}
\bibliography{Moltbook_ArXiv_Submit}

\end{document}